\documentclass[prd,12pt,tightenlines,raggedbottom,titlepage]{revtex4-2}
\pdfoutput=1

\usepackage[usenames,dvipsnames]{xcolor}
\usepackage{hyperref}
\hypersetup{
	colorlinks=true,
	urlcolor=Maroon,
	linkcolor=RoyalBlue,
	citecolor=Maroon,
	bookmarks=true,
	pdftitle={Crossing Symmetry in the Planar Limit},
	pdfauthor={Sebastian Mizera},
	pdfdisplaydoctitle=true,
	pdfstartview=FitH,
	linktocpage=true
}

\usepackage{amsmath}
\usepackage{amsfonts}
\usepackage{amssymb}
\usepackage{mathtools}
\usepackage{microtype}
\usepackage{bm}
\usepackage[utf8]{inputenc}
\usepackage{blkarray}
\usepackage{bigstrut}
\usepackage{nccmath}
\usepackage{enumitem}
\usepackage{braket}
\usepackage{dsfont}
\usepackage[export]{adjustbox}
\usepackage[english]{babel}
\usepackage{physics}
\usepackage{scalerel}
\usepackage[font=small,labelfont=bf,labelsep=period,justification=centerlast,width=.9\textwidth]{caption}
\usepackage[margin=1in]{geometry}
\usepackage{rotating}
\usepackage{tikz}

\usepackage[nodisplayskipstretch]{setspace}
\allowdisplaybreaks
\graphicspath{{figures/}}

\DeclareRobustCommand\lightcone{\begin{tikzpicture}[scale=.2,baseline=-0.3em]
		\draw[thin, black, ->] (-1,-1) -- (1,1) node[scale=0.5, anchor=south west, xshift=-0.1em, yshift=-0.9em] {$+$};
		\draw[thin, black, ->] (1,-1) -- (-1,1) node[scale=0.5, anchor=south east, xshift=-0.1em, yshift=-0.9em] {$-$};
		\draw[densely dotted] (0.6,0.66) arc(0:180:0.6 and 0.2);
		\draw (-0.6,0.66) arc(180:360:0.6 and 0.2);
		\draw[densely dotted] (0.6,-0.66) arc(0:180:0.6 and 0.2);
		\draw (-0.6,-0.66) arc(180:360:0.6 and 0.2);
\end{tikzpicture}}

\def \be {\begin{equation}}
\def \ee {\end{equation}}
\def \nn {\nonumber}
\def \C {\mathbb{C}}

\def \D {\mathrm{D}}
\def \E {\mathrm{E}}
\def \L {\mathrm{L}}
\def \U {\mathcal{U}}
\def \F {\mathcal{F}}
\def \d {\mathrm{d}}
\def \eps {\varepsilon}
\def \V {\mathcal{V}}
\def \G {\mathcal{G}}

\def \leq {\leqslant}
\def \geq {\geqslant}
\def \AA {\mathrm{A}}
\def \BB {\mathrm{B}}
\def \CC {\mathrm{C}}
\def \DD {\mathrm{D}}
\def \II {\mathrm{I}}
\def \JJ {\mathrm{J}}
\def \KK {\mathrm{K}}
\def \LL {\mathrm{L}}

\newcommand{\RN}[1]{\textup{\uppercase\expandafter{\romannumeral#1}}}

\begin{document}

\title{\vspace*{1em}\Large Crossing Symmetry in the Planar Limit}
\author{Sebastian Mizera\\ \footnotesize\href{mailto:smizera@ias.edu}{\texttt{smizera@ias.edu}}}
\affiliation{Institute for Advanced Study, Einstein Drive, Princeton, NJ 08540, USA \vspace{1em}}

\begin{abstract}\linespread{1}\selectfont
Crossing symmetry asserts that particles are indistinguishable from anti-particles traveling back in time. In quantum field theory, this statement translates to the long-standing conjecture that probabilities for observing the two scenarios in a scattering experiment are described by one and the same function. Why could we expect it to be true? In this work we examine this question in a simplified setup and take steps towards illuminating a possible physical interpretation of crossing symmetry. To be more concrete, we consider planar scattering amplitudes involving any number of particles with arbitrary spins and masses to all loop orders in perturbation theory. We show that by deformations of the external momenta one can smoothly interpolate between pairs of crossing channels without encountering singularities or violating mass-shell conditions and momentum conservation. The analytic continuation can be realized using two types of moves. The first one makes use of an $i\varepsilon$ prescription for avoiding singularities near the physical kinematics and allows us to adjust the momenta of the external particles relative to one another within their lightcones. The second, more violent, step involves a rotation of subsets of particle momenta via their complexified lightcones from the future to the past and vice versa. We show that any singularity along such a deformation would have to correspond to two beams of particles scattering off each other. For planar Feynman diagrams, these kinds of singularities are absent because of the particular flow of energies through their propagators. We prescribe a five-step sequence of such moves that combined together proves crossing symmetry for planar scattering amplitudes in perturbation theory, paving a way towards settling this question for more general scattering processes in quantum field theories.
\end{abstract}

\maketitle
\pagestyle{plain}

\tableofcontents

\pagebreak
\section{Introduction}

Classically, particles are indistinguishable from anti-particles with the opposite energy and momentum \cite{Stueckelberg1941}. In order to convert this statement into an observable in quantum field theory, we can phrase it as measuring the two scenarios in a scattering experiment.
At this level, \emph{crossing symmetry} states that on-shell scattering amplitudes for processes involving the particle and the anti-particle are boundary values of one and the same function, regardless of the number and type of the remaining particles it interacts with. It is a fundamentally Lorentzian notion thought to be a reflection of the compatibility of quantum theory with physical principles such as causality, locality, or unitarity.

How are we supposed to think about crossing symmetry? Two scattering processes---one involving an incoming and one with an outgoing state---cannot be directly compared because they are defined in disjoint regions of the momentum space: supported in the future and past lightcones respectively. Therefore, in order to talk about crossing symmetry, one needs to promote the scattering amplitude to a function of complexified kinematics. It is at this stage that the S-matrix becomes a multi-valued function with a host of new singularities---historically referred to as the \emph{anomalous thresholds}---corresponding to possible saddle points of the path integral for a given process. To show that the two scattering amplitudes can be described with a single analytic function, one needs to understand how to navigate around such singularities in order to connect the future and the past lightcones together.

Since its introduction in 1954 by Gell-Mann, Goldberger, and Thirring \cite{GellMann:1954db}, establishing crossing symmetry as a consequence of the aforementioned principles has remained an open problem. In the absence of a convincing physical explanation, the throughline in the literature has been the application of the apparatus of complex analysis in multiple variables, using which one could hope to at least settle the question in the simplest cases. Among these attempts, the most fruitful approach has been undertaken by Bros, Epstein, and Glaser \cite{Bros:1964iho,Bros:1965kbd,Bros:1972jh,Bros:1985gy}, who studied it in the framework of the axiomatic quantum field theory in the Lehmann--Symanzik--Zimmerman (LSZ) formalism.

As with any LSZ-based approach, one is faced with an immediate obstruction stemming from a rather prosaic but highly-consequential fact that Fourier transforms from the position to the momentum space---such as those at the heart of the LSZ formalism---generically do not converge on-shell \cite{Steinmann1960a,Steinmann1960b,ruelle1961connection,doi:10.1063/1.1703695,araki1960properties}. At this stage, one is forced to study \emph{off-shell} Green's functions first and then inferring properties of scattering amplitudes by extensive use of analytic completion theorems in the on-shell limit. (It is not clear if crossing symmetry would be a physically meaningful notion for Green's functions, particularly in the theories with gauge degrees of freedom, perturbative gravity, or those with spontaneously broken symmetries.) The process of analytic continuation is rather abstract and difficult to interpret in terms of particle scattering. In fact, it not only creates an obstruction to understanding the physical origin of crossing symmetry, but also turns out to be technically strenuous. As a consequence, proofs of this type have been completed only in theories without massless particles in the case of $2\to 2$ \cite{Bros:1964iho,Bros:1965kbd} and $2\to 3$ scattering \cite{Bros:1972jh,Bros:1985gy}. For reviews, see \cite{Epstein:1966yea,Sommer:1970mr,bogolubov1989general,10.1007/3-540-09964-6_319}. A more detailed explanation and some clarifications are provided in App.~\ref{sec:appendix}.

The simplest qualitatively distinct cases where crossing symmetry is not known to hold non-perturbatively include: crossing between scattering amplitudes with different numbers of incoming and outgoing states, for example between $2\to 3$ and $3\to 2$ processes; scattering of more than five particles, where constraints from the space-time dimensionality first start to matter; as well as crossing for any process involving massless particles. The lack of a clear physical picture in the aforementioned proofs makes it difficult to ascertain whether crossing symmetry extends to these situations. (There certainly are cases where it does \emph{not} hold, including integrable theories in two dimensions \cite{Iagolnitzer:1977sw} or monopole scattering \cite{PhysRevD.6.458}.)

At this stage we are faced with both conceptual and technical problems. This suggests that a qualitatively new strategy is needed, not only to understand crossing symmetry itself but also shed some new light on the Lorentzian aspects of observables in quantum field theories.
The purpose of this work is to reexamine this problem in a simplified setup. It will allow us to catch a glimpse of what possible meaning could be attached to crossing symmetry, by first identifying all the potential singularities of the S-matrix that could prevent it, and then explaining why they do not exist.

We will be working in the framework of perturbation theory, where one might reasonably hope to address the aforementioned shortcomings.
The main advantage of this approach is that the positions of singularities can be determined by a set of algebraic conditions called \emph{Landau equations} \cite{Landau:1959fi} and have a straightforward interpretation as intermediate particles going on-shell. (Sec.~\ref{sec:Landau} reviews aspects of Landau equations needed for this work.)
In addition, in perturbation theory, the overall divergences and ways of dealing with them are reasonably well-understood with tools such as renormalization or regularization \cite{speer1969generalized,smirnov1991renormalization,zavialov2012renormalized}.
In contrast with the non-perturbative approaches, where amplitudes at different multiplicities had to be analyzed separately, Feynman rules are the same for any number of external states which hints that once the solution to crossing symmetry has been understood at four-point, a similar underlying principle would carry over to more complicated processes.

There have been several attempts at demonstrating crossing symmetry in perturbation theory over the years; see, e.g., \cite{Eden:1966dnq,Iagolnitzer:1978qv,Mizera:2021ujs}. For instance, it is known that any Feynman integral is crossing-symmetric in a large enough space-time dimension, provided its masses satisfy certain bounds \cite{Mizera:2021ujs}. The additional simplification we will make in this paper is to focus on \emph{planar} Feynman diagrams. Nevertheless, we will still consider any multiplicity, masses, spins, and work to all loop orders. For example, we could apply our techniques to the large-$\mathrm{N}$ limit of quantum chromodynamics or other toy-model theories. (Recall that planarity does not mean scattering takes place in a plane, but rather that each Feynman diagram has a planar embedding once all the external lines are extended to infinity.) Although the final step in the proof of crossing will require planarity, the majority of the results described in this work apply to non-planar diagrams.

Other than the intrinsic interest in crossing symmetry as a possible property of quantum field theories, understanding the singularity structure of scattering amplitudes is important in various ``bootstrap'' approaches to the S-matrix theory, which try to constrain the space of allowed observables based on a set of underlying assumptions.
Since the 1960s, it has been customary to take crossing symmetry as an assumption, or even replace it with much stronger conditions on analyticity, which at various levels state that scattering amplitudes are non-singular away from the physical regions.
While there is nothing wrong with making simplifying assumptions, in view of the author, axioms without a clear physical meaning should not form a basis for a physical theory.

\subsection{Summary of the Analytic Continuation}

We focus on planar scattering amplitudes preserving a given cyclic ordering $(\AA\BB\CC\DD)$ of $n$ external states, where each of $\AA$, $\BB$, $\CC$, and $\DD$ denote non-empty sets of particles. Let us label the incoming particles with $\AA\BB$ and the outgoing ones with $\CC\DD$.
Provided that the connected part of the scattering amplitude ${\cal T}_{\AA\BB \to \CC\DD}$ exists, we will show that it can be analytically continued to the crossed process:
\be\label{eq:AB-CD}
{\cal T}_{\AA\BB \to \CC\DD} \;=\; {\cal T}_{\BB\bar{\CC} \to \DD\bar{\AA}},
\ee
where the bar denotes changing particles into anti-particles.
To be precise, the meaning of the equality sign above is that there exists a complex analytic function whose boundary values in their respective physical regions are the two scattering amplitudes. Description of the precise path of the analytic continuation is the content of this paper. Note that we will not consider crossing between channels in which the incoming and outgoing sets are not consecutive, unless they appear in the intermediate steps outlined below. Likewise, the requirement of each set $\AA$, $\BB$, $\CC$, and $\DD$ being non-empty excludes the case of particle decay, which has to be treated separately when it is kinematically allowed. The challenge is to perform the analytic continuation in such a way that the momentum conservation and the mass-shell conditions are not violated along the path of deformation.

Let us remind the reader that crossing symmetry should not be confused with permutation invariance (or cyclic invariance for planar diagrams), which only means that whenever two identical states are exchanged, their scattering amplitudes are relabelings of each other, up to spin-statistics, but are nevertheless defined in disjoint physical regions. Likewise, CPT invariance is a prerequisite for crossing symmetry, but---being a kinematic not a dynamic statement---by no means guarantees it.

By repeated use of \eqref{eq:AB-CD} we can analytically continue between processes involving a particle and an anti-particle without affecting the remaining states. To see this, let us start with a process $\II\JJ \to \KK\LL n$ for some non-empty sets $\II$, $\JJ$, $\KK$, and $\LL$. We can convert the particle $n$ into an anti-particle $\bar{n}$ with the following steps: 
\be
{\cal T}_{\,\II\JJ \to \KK\LL n} \;=\;
{\cal T}_{\JJ\bar{\KK}\bar{\LL} \to n\bar{\II}} \;=\;
{\cal T}_{\,\bar{\LL}\bar{n} \to \bar{\II}\bar{\JJ}\KK} \;=\;
{\cal T}_{\bar{n}\II\JJ \to \KK\LL}.
\ee
Here each equality is a special case of \eqref{eq:AB-CD}: in the first step $\BB=\JJ$, $\CC=\KK\LL$; in the second one $\BB=\bar{\LL}$, $\CC=n$; and finally in the third $\BB = \bar{n}$, $\CC=\bar{\II}\bar{\JJ}$. As a result, we obtained the scattering amplitude for the crossed process $\bar{n}\II\JJ \to \KK \LL$, where only $\bar{n}$ changed its nature and all the other states remained untouched.
It is the first class of scattering processes in which crossing symmetry, under the definition given above, can be directly demonstrated.

\begin{figure}[!t]
	\includegraphics[scale=1.1]{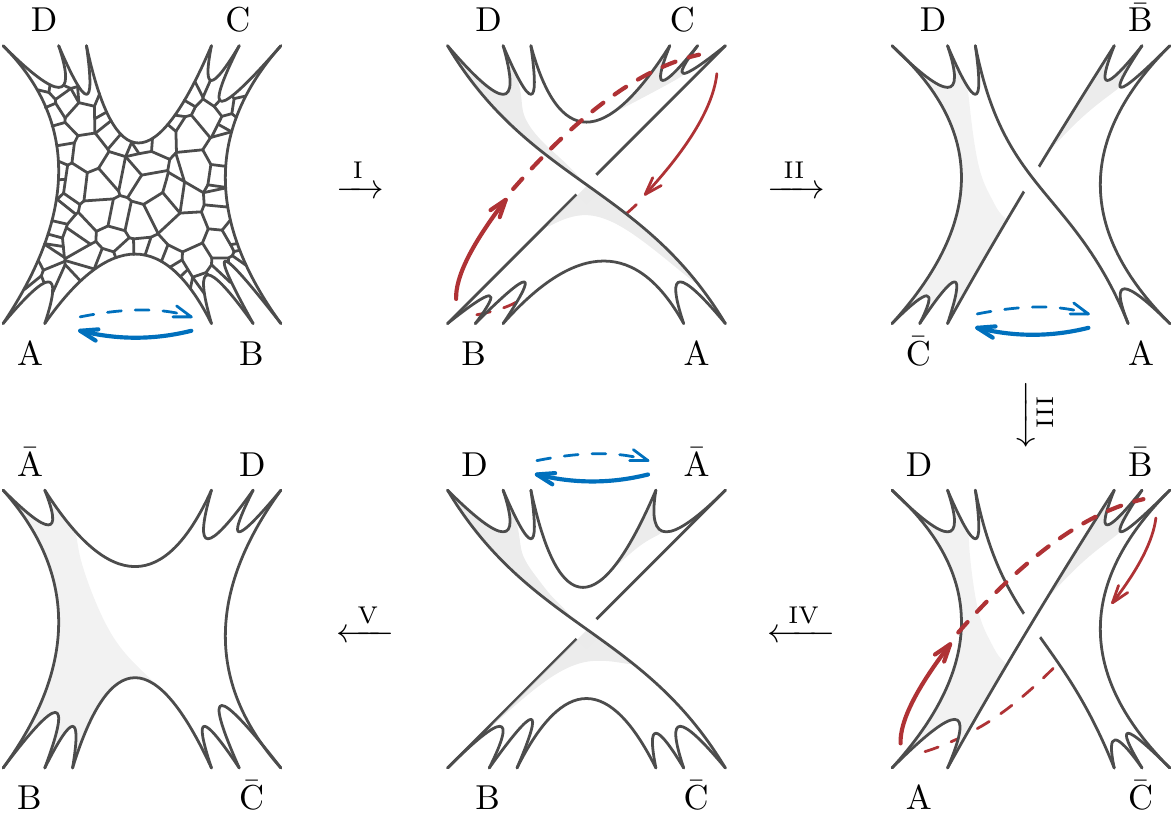}
	\caption{\label{fig:steps}Cartoon depiction of the analytic continuation of a planar Feynman diagram between the crossing channels $\AA\BB \to \CC\DD$ and $\BB\bar{\CC} \to \DD\bar{\AA}$. The pictures are meant to be embedded in a lightcone \!\!\lightcone with time directed upwards. For readability we do not draw internal propagators after the first step and simply represent the planar diagram as a surface. The individual steps are described in the main text.}
\end{figure}

The analytic continuation behind \eqref{eq:AB-CD} is illustrated on a cartoon level in Fig.~\ref{fig:steps}. Since we would like to understand it at the level of particle momenta---as opposed to the Mandelstam invariants---we first make a choice of a Lorentz frame.
We then prescribe a specific deformation of the external momenta in such a way that the amplitude navigates around any possible singularities. Recall that singularities (or anomalous thresholds) can develop when a subset of propagators goes on-shell, and accordingly the diagrams in Fig.~\ref{fig:steps} are meant to depict precisely the configurations that can never go on-shell for any real value of the loop momenta.

The deformation proceeds in five steps, but there are only two distinct moves: in the first one the momenta stay in their original lightcones (blue), and in the second one they are rotated from the future to the past lightcones and vice versa (red).

Step $\RN{1}$ aligns the momenta within their lightcones in such a way that the sets $\BB$ and $\CC$ lie closer to its positive axis (at the $\tfrac{\pi}{4}$ angle) than the sets $\AA$ and $\DD$ respectively. Along such a deformation we might encounter singularities, but they can be easily avoided using an appropriate $i\eps$ deformation in the kinematic invariants, related to the Feynman $i\eps$ prescription and preservation of causality. We describe this procedure in detail in Sec.~\ref{sec:physical}.

In step $\RN{2}$ we simultaneously rotate $\BB$ and $\CC$ through their complexified lightcones in such a way that signs of their energies get exchanged, resulting in the sets of anti-particles $\bar{\BB}$ and $\bar{\CC}$. We call the region of the kinematic space through which such a continuation takes place a \emph{crossing domain}.
Along this deformation we can look at the scattering process from the real and the imaginary directions. Since only a subset of the momenta are complexified, in the imaginary directions it looks \emph{as if} it corresponded to a highly-energetic process---with all the energies large, while all the transfer momenta and masses are vanishingly small---even though in reality all the Mandelstam invariants stay finite. We will show that an on-shell singularity for such a process can only develop if it looks like two beams of particles (with all the internal momenta aligned either in the $\BB$--$\CC$ or in the $\AA$--$\DD$ directions) scattering off each other. This is a completely general statement, since up to this stage we made no use of planarity. Therefore, proving crossing symmetry amounts to explaining why such singularities cannot appear.

In order to show analyticity in the crossing domains, we make use of the peculiar nature of the energy flow in planar diagrams, as explained in Sec.~\ref{sec:energy-flow}. To be more precise, we will show that propagators along the sides of a generic planar diagram must always have a non-zero energy component in \emph{both} the $\BB$--$\CC$ and $\AA$--$\DD$ directions at the same time. This fact is powerful enough to prove that no singularities can appear along the precise path of analytic continuation between the future and past lightcones prescribed in Sec~\ref{sec:crossing}.

The remaining steps use essentially the same arguments up to permutations of labels. Step $\RN{3}$ is analogous to step $\RN{1}$ but adjusts $\AA$ and $\bar{\BB}$ to lie closer to the positive axis of the lightcone than the remaining sets of particles. In step $\RN{4}$ we once again rotate $\AA$ and $\bar{\BB}$ through the complex directions such that they become $\bar{\AA}$ and $\BB$ respectively. Finally, in step $\RN{5}$ we continue to a process with generic momenta, avoiding any singularities using the $i\eps$ prescription just like in step $\RN{1}$. As a result, we end up with the scattering amplitude ``rotated'' as in Fig.~\ref{fig:steps}, which is now defined in the $\BB\bar{\CC} \to \DD\bar{\AA}$ physical region.

\subsection{Four-Point Example}

\begin{figure}[!t]
	\includegraphics[scale=1.1]{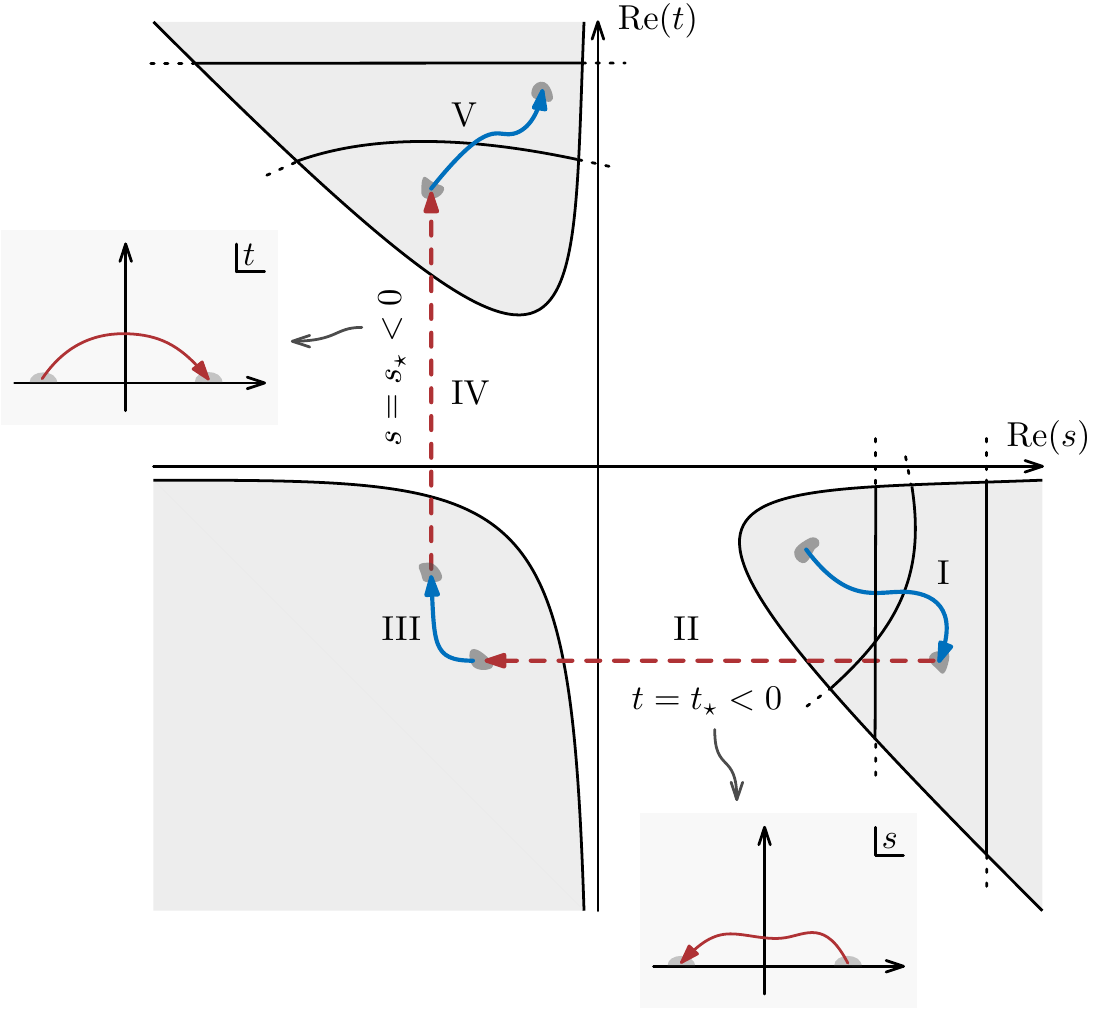}
	\caption{\label{fig:ResRet}Analytic continuation in the $\Re(s)$--$\Re(t)$ plane for $n=4$. Dashed lines in steps $\RN{2}$ and $\RN{4}$ denote continuation in the upper-half $s$- and $t$-planes respectively.}
\end{figure}

In order to illustrate the Lorentz-invariant content of this procedure, let us exemplify it in the four-point case. Here we have two independent Mandelstam invariants: the center of mass energy $s$ and the squared momentum transfer $t$. The path of analytic continuation is schematically illustrated in Fig.~\ref{fig:ResRet}. Before we go through this path, let us emphasize that in the case $n=4$ specifically, an analytic continuation for planar Feynman diagrams can be performed more easily. We nonetheless use the specific procedure outlined in Fig.~\ref{fig:steps}, because it is the one that generalizes to arbitrary multiplicity $n$ and also has an interpretation in terms of the momentum vectors. The essential difficulty with prescribing analytic deformation directly at the level of the Mandelstam invariants is that for $n>5$ one would have to deal with complications related to constraints coming from the dimensionality of space-time.
Finally, since we are ultimately interested in the non-planar cases, we want to avoid hard-wiring planarity as much as possible.

In this example the goal is to analytically continue between physical regions (shaded) in the $s$-channel (bottom right) and the $t$-channel (top left). We assume that the scattering amplitude exists in the first place and hence can be defined in some open set around the starting point of step $\RN{1}$. Adjusting the momenta within their lightcones leads to a blue path moving within the $s$-channel physical region. Since it might potentially encounter some singularities (black), we prescribe an $i\eps$ rule for going around them in the direction consistent with causality. (The unshaded regions also contain a myriad of singularities not illustrated in the figure, since they do not concern us.) The image of step $\RN{2}$ in the kinematic invariants corresponds to keeping $t$ fixed to some $t_\star<0$ and continuing in the upper-half plane of $s$. It ends up in the $u$-channel region (bottom left). For a non-planar process, this way of approaching the $u$-channel would be unphysical, which is why it is taken only as an auxiliary step in the analytic continuation. The remaining steps are identical up to relabeling $s \leftrightarrow t$.

The above procedure \emph{defines} the ``physical sheet'' in the space of Mandelstam invariants, along which analytic continuation between different physical regions can be made. Recall that such a notion would not be meaningful unless crossing symmetry is proven.

In contrast with the non-perturbative results \cite{Bros:1965kbd}, which require continuations to asymptotically large $|s|$ and $|t|$ at intermediate steps, the domain of analyticity described above is substantially larger. Although due to the aforementioned issues with the physical interpretation of the non-perturbative results, it is difficult to pinpoint the origin of this difference, we strongly suspect it should be associated with the simplification of planarity rather than that of perturbation theory.

\pagebreak
\vspace{1em}\noindent{\bf Outline.} In Sec.~\ref{sec:Landau} we start by reviewing different formulations of Landau equations, their meaning in the worldline formalism, and what is known about their solutions.
In Sec.~\ref{sec:physical} we show analyticity in the neighborhoods of the physical regions using small $i\eps$ deformations of the kinematic invariants and the integration variables. In Sec.~\ref{sec:crossing} we demonstrate analyticity in the crossing domains from the perspective of the loop momenta and the worldline formalism. We finish with a brief discussion of the future directions in Sec.~\ref{sec:outlook}. In App.~\ref{sec:appendix} we provide some clarifications regarding Fourier transforms of retarded commutators.

\vspace{1em}\noindent{\bf Conventions.} Incoming momenta are called $p_i^\mu$ and the outgoing ones $-p_i^\mu$, such that scattering amplitudes of $n$ particles are supported on the constraints of the momentum conservation $\sum_{i=1}^{n} p_i^\mu = 0$ and the mass-shell conditions $p_i^2 = M_i^2 \geq 0$ in the mostly-minus signature. We work in asymptotically-flat space-time with dimension $\D$. Although the main physical interest lies in $\D=4$, the arguments work in any fixed $\D > 2$.

\section{\label{sec:Landau}Review of Landau Equations}

We will consider a perturbative expansion of scattering amplitudes in terms of Feynman diagrams.
Feynman rules of a local and CPT-invariant quantum field theory already guarantee that each Feynman \emph{diagram} has the expected crossing property. Therefore, the crux of the problem lies in showing that Feynman \emph{integrals} cannot develop singularities when analytically continued between crossing channels.

At this stage, one should distinguish between two separate types of issues: the overall divergences (such as UV or IR divergences) and singularities (associated to resonances).
The former happen for every value of the external kinematics, while the latter only at specific values of Mandelstam invariants corresponding particles going on-shell.
In order to meaningfully formulate the question of crossing symmetry, we must assume that the amplitude exists in the first place, i.e., it can be defined in some open set of the physical region of interest.
This can be done using standard tools such as the Bogoliubov--Parasiuk--Hepp--Zimmermann (BPHZ) renormalization, or dimensional or analytic regularization (see, e.g., \cite{speer1969generalized,smirnov1991renormalization,zavialov2012renormalized} for reviews), without affecting singularities.
In the worst-case scenario, instead of Feynman integrals one can talk about Feynman integrals in dimensional regularization, which are always free of the overall divergences. From now on we assume that the amplitude exists in one of the above senses. Moreover, we only consider amplitudes defined at generic kinematic points, e.g., not evaluated directly on a collinear limit.

We will consider two complementary ways of looking at Feynman integrals: in the loop-momentum space and in the worldline formalism. 

\subsection{Loop Momentum Perspective}

We start with a scalar Feynman integral with $n$ external legs, $\L$ loops, and $\E$ internal edges (propagators) in $\D$-dimensional Minkowski space, which can be written as
\be\label{eq:I}
{\cal I} := \int \d^{\D \L} \ell_I \prod_{e=1}^{\E} \frac{i \hbar}{q_e^2 - m_e^2 + i\eps},
\ee
where the product runs over all the propagators. Here $q_e^\mu$ and $m_e$ are the momenta and masses associated to the edge $e$ and $\ell_I^\mu$ denote the loop momenta. They satisfy momentum conservation at every vertex $v$,
\be\label{eq:LE1}
p_v^\mu + \sum_{e=1}^{\E} \eta_{ve} q_e^\mu = 0,
\ee
with $p_v^\mu$ denoting the total external momentum flowing into this vertex, which we assume to be real. Here $\eta_{ve}$ equals to $+1$ ($-1$) if the edge $e$ is incoming (outgoing) from this vertex and $0$ otherwise. We assign arbitrary orientations to all loops and edges.

In order to analyze singularities of Feynman integrals it is convenient to express the propagators as integrals over Schwinger parameters $\alpha_e$ as follows:
\be
\frac{i \hbar}{q_e^2 - m_e^2 + i\eps} = \int_0^\infty \d \alpha_e\, e^{\frac{i}{\hbar}(q_e^2 - m_e^2 + i\eps) \alpha_e}.
\ee
Note that from this perspective the Feynman $i\eps$ is needed to ensure convergence near infinity. Whenever $i\eps$ appears, such as in \eqref{eq:I}, the expression should be understood as a limit where $\eps \to 0^+$. Applying this procedure $\E$ times one finds
\be\label{eq:I2}
{\cal I} = \int \d^{\E} \alpha_e\, \d^{\D\L} \ell_I\, e^{\frac{i}{\hbar}(\V \,+\, i\eps \sum_{e=1}^{\E}\! \alpha_e)},
\ee
where the integration domains will remain suppressed from the notation from now. Here we defined
\be\label{eq:V}
\V(\alpha_{e}, \ell_I) := \sum_{e=1}^{\E} (q_e^2 - m_e^2)\, \alpha_e.
\ee
This quantity will allow us to translate between different ways of thinking about singularities of Feynman integrals.

\subsubsection{Bulk Saddle Points}

The most obvious singularities are associated to the saddle points in the classical limit, $\hbar \to 0$, inside the integration domain. They are obtained by extremizing $\V$. Doing so with respect to the loop momenta $\ell_I^\mu$ one finds
\be\label{eq:LE2}
\sum_{e=1}^{\E} \eta_{I e} q_e^\mu \alpha_e = 0,
\ee
for each $I=1,2,\ldots,\L$, as a result of the fact that each internal momentum $q_e^\mu$ is linear in the loop momenta $\ell_I^\mu$. Here $\eta_{I e}$ equals to $+1$ $(-1)$ if the orientations of the loop $I$ and the edge $e$ match (mismatch) and $0$ otherwise. Extremizing with respect to the Schwinger parameters $\alpha_e$ yields
\be\label{eq:LE3}
q_e^2 - m_e^2 = 0
\ee
for each $e=1,2,\ldots,\E$.

The equations \eqref{eq:LE1} together with \eqref{eq:LE2} and \eqref{eq:LE3} are known as the \emph{leading Landau equations}. They were introduced in 1959 independently by Bjorken \cite{Bjorken:1959fd}, Landau \cite{Landau:1959fi}, and Nakanishi \cite{10.1143/PTP.22.128,10.1143/PTP.23.284}. Because of their degree in the momenta, we will from now on refer to (\ref{eq:LE1},\ref{eq:LE2}) as the \emph{linear} and to \eqref{eq:LE3} as the \emph{quadratic} Landau equations. The general strategy will be to solve the linear part explicitly and then constrain the solutions of the quadratic part.

Singularities associated with solutions of Landau equations are also called anomalous thresholds, unless the dimension spanned by the internal momenta $q_e^\mu$ is exactly one, in which case they would be referred to as normal thresholds. The physical interpretation of leading Landau equations is that of every internal propagator going on-shell in the classical limit. In other words, the path integral for the scattering process localizes on a specific configuration in which the corresponding Feynman diagram describes a physical interaction of long-lived particles in space-time \cite{Coleman:1965xm}.

Solutions of Landau equations are generically complex in the Mandelstam invariants and the Schwinger parameters. Note that they are projective in the $\alpha_e$'s (for massless theories they are also projective in the external kinematics), which means that if a given set $\alpha_e^\ast$ is a solution, so is $\lambda \alpha_e^\ast$ for any $\lambda \in \C {\setminus} \{0\}$. This means there are more equations than the number of independent Schwinger parameters, resulting in (at least) one constraint on the external kinematics. Additionally, we can distinguish between real projective, $\alpha_e^\ast \in \mathbb{RP}^{\E-1}$, and complex projective, $\alpha_e^\ast \in \mathbb{CP}^{\E-1}$, solutions, out of which only the former contribute directly to the singularities of Feynman integrals on the undeformed integration contours. We will return to this distinction in Sec.~\ref{sec:LE-review}.

\subsubsection{Boundary Saddle Points}

In the analysis so far we have ignored the saddle points confined to the boundaries of integration, which also contribute to the classical limit. We can have boundary saddles corresponding to $\alpha_e \to 0$ (or $\alpha_e \to \infty$) for a subset of Schwinger parameters, as well as $\ell_I^\mu \to \pm \infty$ for a subset of loop momenta. Naturally, we could also encounter mixed types of singularities where multiple of the above criteria are satisfied at the same time.

The first type of singularities are determined by the \emph{subleading} Landau equations, which are identical to the leading ones, but for a simpler diagram with a subset of edges $e$ (or its complement) contracted. In our analysis we will consider all possible diagrams anyway, so without loss of generality we can focus on the leading Landau equations with $\alpha_e > 0$ from now on. (It should also be pointed out that edge contractions preserve planarity.) In the literature, a Feynman diagram with a subset of propagators on-shell is called a \emph{reduced} or \emph{on-shell diagram}.

The boundary saddle points corresponding to infinite loop momenta are called \emph{second-type} Landau singularities \cite{Cutkosky:1960sp,doi:10.1063/1.1724262,Drummond1963}. As will become clear in the next subsection, they do not need to concern us either.

\subsection{Worldline Perspective}

In the representation \eqref{eq:I2}, the argument of the exponent is quadratic in the loop momenta $\ell_I^\mu$. We can therefore simply integrate them out. This gives us
\be\label{eq:I3}
{\cal I} = c\! \int \frac{\d^\E \alpha_e}{\U^{\D/2}}\, e^{\frac{i}{\hbar}(\V + i\eps \sum_{e=1}^{\E}\alpha_e )} ,
\ee
where $c$ is a constant that does not matter for the question of singularities. Here we used the same symbol for the exponent as in \eqref{eq:V}, because on the Gaussian saddle point they are given in terms of the same function,
\be\label{eq:V3}
\V(\alpha_e) = \sum_{e=1}^{\E} (q_e^2 - m_e^2)\, \alpha_e \Big|_{(\ref{eq:LE1},\ref{eq:LE2})}.
\ee
It is now a function of the Schwinger parameters and Mandelstam invariants only. We will give an explicit solution to the linear Landau equations (\ref{eq:LE1},\ref{eq:LE2}) shortly. Additionally, the factor $\U$ can be expressed in terms of a sum over all possible spanning trees $T$,
\be\label{eq:U}
\U := \sum_{\substack{\text{spanning}\\ \text{trees }T}} \prod_{e' \notin T} \alpha_{e'},
\ee
where each term is weighted with the Schwinger parameters of the $\L$ edges needed to be removed from the original diagram to give the tree $T$. It is a homogeneous polynomial with degree $\L$ in the $\alpha_e$'s. On the undistorted integration contour we have $\U > 0$.

The representation \eqref{eq:I3} has a natural interpretation descending from the worldline formalism, where $\V$ denotes the action after localization of the worldline fields and $\U$ is the determinant of the reduced Laplacian. For this reason we will refer to $\V$ as the \emph{action} from now on. In fact, this picture clarifies the meaning of the integration domain of \eqref{eq:I3} as that over the equivalence classes of Riemannian metrics on the Feynman graph, which eventually becomes important in accounting for nested divergences of Feynman integrals, though we do not consider them here. Similarly, in this language the equations \eqref{eq:LE1} and \eqref{eq:LE2} are the Gauss constraint and the continuity law for the worldline fields respectively. The projective redundancy $\alpha_e \sim \lambda\alpha_e$ descends from the reparameterization invariance of the worldlines.

There exist special kinematic configurations in which one cannot perform the Gaussian integral in \eqref{eq:I2} to obtain \eqref{eq:I3}. They appear precisely when a finite solution of the linear Landau equations (\ref{eq:LE1},\ref{eq:LE2}) does not exist, or equivalently $\U=0$ (see also the explicit solution \eqref{eq:qe} later on). These are the aforementioned Landau singularities of the second type.
They only exist when the external momenta are collinear, since the system of equations (\ref{eq:LE1},\ref{eq:LE2}) can degenerate only if there is a linear constraint on the external data $p_i^\mu$ beyond the momentum conservation.
The reason why we do not need to consider such singularities is not only that they appear at special kinematic configurations, but also that they happen only after deformation of the contour, e.g., when computing singularities of discontinuities, but not the amplitude itself \footnote{Similarly, in integrating out the loop momenta in \eqref{eq:I3}, we have explicitly factored out the contributions from the saddle points purely in the loop-momentum directions, i.e., those that are valid for all the values of Schwinger parameters $\alpha_e$. The simplest example is the origin of the kinematic space for massless processes, where Landau equations are trivially satisfied for all $\alpha_e$'s: one of the early signs that in massless theories the kinematic space should be thought of in projective terms.}.

At this stage let us point out that the interpretation of Landau equations as (stratified) saddle points of the worldline action is not standard in the literature, but we believe it gives the most intuition for what Landau equations mean physically \footnote{However, one should remember that because of the homogeneity of $\V$, in reality the Feynman integral ${\cal I}$ has only an overall $\hbar$-dependence, as can be seen from \eqref{eq:I}.}. Rigorous formulation can be achieved with stratified Morse theory.

\subsubsection{Spins and Numerators}

We are now equipped to comment on the meaning of Landau equations for theories with spin. In either of the representations (\ref{eq:I},\ref{eq:I2},\ref{eq:I3}), spin effects can at most multiply the integrand by a polynomial in the integration variables; see, e.g., \cite{smirnov1991renormalization} for details. They can therefore remove or change the nature of on-shell singularities, but never introduce new ones. In other words, Landau equations give necessary but not sufficient conditions for the development of singularities. They depend on the topology of the diagram and kinematics, but not on the specificities of the interaction vertices. 
In general it is difficult to determine a priori if a given numerator cancels a singularity, which can happen in highly non-trivial ways, for example in theories with dual conformal invariance. (Under certain restrictions on the numerators, one can argue that Landau equations are not only necessary but also sufficient conditions for singularities \cite{Collins:2020euz}.) In particular, although Landau equations do not explicitly depend on $\D$, the presence of singularities does.
Nevertheless, in our applications we aim to show that even the necessary conditions cannot be satisfied along a path of analytic continuation and hence the discussion applies to any spin.

\subsubsection{Causality}

At the level of the representation \eqref{eq:I3}, the individual $i\eps$ factors lost their original meaning and instead the correct causality conditions are imposed by requiring that the integrand of \eqref{eq:I3} decays sufficiently fast at infinities. Instead of introducing the $i\eps$ ``by hand,'' in the following we will instead deform the external kinematics (with momentum conservation and on-shell conditions satisfied) as well as the Schwinger parameters such that

\pagebreak
\be
\Im \V > 0,
\ee
and approach the physical regions with $\Im \V \to 0^+$. To be more precise, this prescription selects the correct homology class of integration contours giving rise to physical amplitudes. Since we have to deform the external kinematics for the purposes of the analytic continuation in any case, in this way we can avoid the introduction of an additional variable and the associated tedious discussion of the order of limits.

\subsubsection{Expressions for the Action}

The action $\V$ has multiple different representations that come useful in various applications \cite{10.1143/PTPS.18.1,nakanishi1971graph,Pohlmeyer:1974ph,Trute:1974nt,Bogner:2010kv,Mizera:2021ujs}. Already from \eqref{eq:V} it is clear that $\V$ has to be linear in Mandelstam invariants and masses. Particularly useful for us will be the form
\be\label{eq:V2}
\V(\alpha_e) = \sum_{\text{subsets }S}\!\! p_S^2\, {\cal F}_S - \sum_{e=1}^{\E} m_e^2 \alpha_e,
\ee
where the first sum runs over all the $2^{n-1}-1$ proper subsets $S$ of $n$ labels without double-counting the complements $\bar{S} := \{1,2,\ldots,n\} \setminus S$ with
\be
p_S^\mu := \sum_{i \in S} p_i^\mu.
\ee
The function $\F_S$ is defined through
\be\label{eq:FS}
\F_S := \frac{1}{\U} \sum_{\substack{\text{spanning}\\ \text{2-forests }F_{S}}} \prod_{e \notin F_S} \alpha_e,
\ee
where a spanning $2$-forest $F_{S} := T_S \sqcup T_{\bar{S}}$ is a disjoint union of trees $T_S$ ($T_{\bar{S}}$) connected to all the external momenta from the set $S$ ($\bar{S}$) and none from $\bar{S}$ ($S$) such that every vertex belongs to either tree. This can be achieved by cutting through exactly $\L{+}1$ edges such that the sets $S$ and $\bar{S}$ are separated, and hence every $\F_S$ is a homogeneous function with degree one in the $\alpha_e$'s.
In the literature $\U$ is called the first Symanzik polynomial and $\V$ is the ratio of the second to the first Symanzik polynomials.

The advantage of the form \eqref{eq:V2} is that it drastically simplifies for planar diagrams. This is because such diagrams depend only on $n(n{-}3)/2$ Mandelstam invariants of the form $p_S^2$ where $S$ are sets of consecutive labels with respect to a given planar ordering (here the $n$ masses $p_i^2 = M_i^2$ are not counted). When the number of external particles is $n \leq \D{+}1$, all the Mandelstam invariants can be independently deformed. At higher multiplicity, $n > \D{+}1$, this cannot be done in a simple manner because of the additional Gram matrix constraints on the kinematics. The approach in the later sections will be to deform the four-momenta $p_i^\mu$ directly, which circumvents this issue.

\subsubsection{Singularities}

In this language the leading Landau equations are simply the saddle-point conditions
\be
\frac{\partial \V}{\partial \alpha_e}  = q_e^2 - m_e^2\,\Big|_{(\ref{eq:LE1},\ref{eq:LE2})} = 0
\ee
for every $e = 1,2,\ldots,\E$. They retain their previous meaning of propagators going on-shell, essentially because the saddle points in the loop-momentum directions were Gaussian. Since $\V$ is homogeneous with degree one, on the saddle point we also have
\be\label{eq:V0}
\V = \sum_{e=1}^{\E} \alpha_e \frac{\partial \V}{\partial \alpha_e} = 0,
\ee
which itself is a necessary, but not a sufficient, condition for a singularity. For example, whenever $\V=0$ but not all derivatives of $\V$ vanish, one can deform the integration contour to avoid such a situation, which happens, e.g., on the threshold cuts. As we will see in Sec.~\ref{sec:non-singular}, this cannot be done on the saddle points.

Singularities arise because on the support of \eqref{eq:V0} there is a projective family of integrands no longer suppressed in the $\eps \to 0^+$ limit.
To be specific, under the rescaling $\alpha_e \mapsto \lambda\alpha_e$ we have
\be
\U \,\mapsto\, \lambda^{\L}\, \U,\qquad \V \,\mapsto\, \lambda \V, \qquad \d^\E \alpha_e \,\mapsto\, \frac{\d^{\E}\alpha_e}{\mathrm{GL}(1)} \frac{\d\lambda}{\lambda^{1-\E}},
\ee
because of their homogeneity properties. The symbol $\mathrm{GL}(1)$ in the denominator indicates a quotient by the overall scale $\lambda$, which is typically fixed by imposing $\sum_{e=1}^{\E}\alpha_e = 1$ or $\alpha_{e'}=1$ for a single edge $e'$. Therefore, on the projective solution of the Landau equations $\alpha_e^\ast \sim \lambda \alpha_e^\ast$, the $\lambda$-dependent part of the integrand is proportional to
\be
\int_{0}^{\infty} \frac{\d \lambda}{\lambda^{1-\gamma}} e^{\frac{i\lambda}{\hbar} (\V^\ast + i\eps \sum_{e=1}^{\E} \!\alpha_e^\ast)} \;\propto\;  \frac{\Gamma(\gamma)}{(\V^\ast + i\eps\, \medmath{\textstyle\sum_{e=1}^{\E}\! \alpha_e^\ast})^\gamma}.
\ee
Here $\gamma = \E - \L\D/2$ is the superficial degree of divergence. Even in dimensional regularization, the integral diverges when $\gamma \geq 0$ since $\V^\ast = 0$ on the saddle point. Since for our purposes we are interested in proving that even the necessary conditions for singularities cannot be satisfied, we will not distinguish between different signs of $\gamma$ in the following.

For completeness let us mention that on real kinematics and away from Landau singularities, integrating out the overall scale leads to the textbook representation of Feynman integrals
\be\label{eq:I4}
{\cal I} = c'\, \Gamma(\gamma)\! \int \frac{\d^\E \alpha_e}{\mathrm{GL}(1)}\, \frac{1}{\U^{\D/2}\, (\V + i\eps\, \medmath{\textstyle\sum_{e=1}^{\E} \!\alpha_e})^\gamma},
\ee
where $c'$ is a constant. From this perspective, Landau equations give necessary conditions for pinch singularities, which in the simplest case corresponds to two zeros of $\V$ trapping the integration contour. We will not be using this point of view.

\subsection{\label{sec:LE-review}What is Known About Solutions of Landau Equations?}

To help the reader maneuver through the literature, in this subsection we pause briefly to review what is known about solutions of Landau equations. Alas, this interlude will not be long, reflecting the fact this topic has been understood rather poorly. There have been multiple textbooks written on this subject, e.g., from the perspective of combinatorics \cite{nakanishi1971graph}, complex analysis \cite{Eden:1966dnq,todorov2014analytic}, algebraic geometry and topology \cite{Hwa:102287,pham1967introduction,pham2011singularities}, and axiomatic quantum field theory \cite{Iagolnitzer:1994xv}.

Let us briefly come back to the distinction between solutions of Landau equations with real positive and complex Schwinger parameters.
The first type of solutions, also known as positive-$\alpha$ or $+\alpha$-Landau surfaces, has a special meaning, because it corresponds to the singularities on the undistorted integration contour. They have an interpretation as classical particles propagating in space-time where---at least in the massive cases---the Schwinger parameters are proportional to the proper time of the corresponding particle \cite{Coleman:1965xm}. In contrast with complex solutions, the real ones have been extensively studied and are reasonably well-understood, particularly due to the work of Chandler, Iagolnitzer, and Stapp, e.g., from the perspective of macroscopic causality and microanalyticity \cite{PhysRev.174.1749,doi:10.1063/1.1664486,Iagolnitzer:1968zz,Iagolnitzer1969,Chandler:1969bd,doi:10.1063/1.1664876}, cuts and discontinuities \cite{doi:10.1063/1.1704822,Coster:1970jy,Cahill:1972ye,1977155}, or Steinmann relations \cite{AIHPA_1967__6_2_89_0,Boyling1968,CAHILL1975438,doi:10.1063/1.522682}. Complex solutions are equally important when large (not infinitesimal) contour deformations are needed, for example in order to determine singularities of discontinuities, but no general criteria exist to straightforwardly tell if they actually contribute (see, e.g., \cite{ANDERSSON1966501,doi:10.1063/1.1703855} for partial progress).

How special are the real solutions? One answer to this question can be obtained by considering the Landau singularity in a single Mandelstam invariant, say $s = s^\ast(m_e,\alpha_e^\ast)$, as a function of the internal masses $m_e$ and the Schwinger parameters $\alpha_e^\ast$ with the remaining Mandelstam invariants fixed. Varying \eqref{eq:V2} with respect to all possible $m_e^2$ gives
\be
\Im \left(\alpha_e^\ast \middle/ \alpha_{e'}^\ast \right) = \Im \left(\frac{\partial s^\ast}{\partial m_e^2} \middle/ \frac{\partial s^\ast}{\partial m_{e'}^2}\right)
\ee	
for any pair $e$ and $e'$. The right-hand side is nothing but the condition for the envelope of a family of Landau curves as the variables $m_e^2$ vary in the $s$-plane. We conclude that ratios of Schwinger parameters are only real on such envelopes, or at the boundary of variation (in the massless limit $m_e = 0$) \cite{Kallen:1961wza,ANDERSSON1965601}. Complex solutions therefore appear generically even for real kinematics, as one can verify already in the simplest examples \cite{doi:10.1063/1.1703752,doi:10.1063/1.1703897,RevModPhys.36.833}.

Explicit solutions of Landau equations are limited to diagrams with small numbers of loops and legs, with at least partial results up to $n \leq 5$; see, e.g., \cite{Olive1962,doi:10.1063/1.1704978,doi:10.1063/1.1664557}. In order to make progress for general diagrams, one can employ a strategy called \emph{majorization}, which shows that when all the masses are equal, an arbitrary Feynman diagram cannot have worse singularities than a finite set of diagrams in certain region near the origin of the real kinematic space (including the so-called ``Euclidean'' region), allowing to prove results such as dispersion relations \cite{10.1143/PTP.20.690,Wu:1961zz}. These techniques have been recently streamlined and extended to non-equal masses at arbitrary multiplicity \cite{Mizera:2021ujs} in the context of crossing symmetry.

Large simplifications come with massless and planar diagrams, which acquired interest in the context of ${\cal N}=4$ super Yang--Mills theory, where Landau equations can be formulated in the momentum twistor space; see, e.g., \cite{Dennen:2015bet,Dennen:2016mdk,Prlina:2017azl,Prlina:2017tvx,Gurdogan:2020tip}. In such cases one can show that at fixed $n$, Landau singularities for an arbitrary-loop Feynman diagram can be no worse than that of a single \emph{ziggurat} diagram \cite{Prlina:2018ukf}. For Feynman integrals enjoying dual conformal invariance---where the number of kinematic variables drops drastically---the explicit solution has been found for $n=6$ \cite{Prlina:2018ukf}.

Rigorous study of Landau singularities has been pioneered by Pham \cite{AIHPA_1967__6_2_89_0,pham1967introduction} and collaborators, in particular with the applications of the ambient isotopy theorem \cite{FOTIADI1965159,AnalyticStudy1,AnalyticStudy2}. They are often formulated in terms of a potential function on the cotangent bundle of Minkowski space.
Landau equations turn out to hide rich geometric structure, whose study has been undertaken, e.g., from the perspective of
the theory of hyperfunctions and holonomic systems \cite{10.1007/BFb0062917,10.1007/BFb0013296,Kashiwara1979};
monodromy groups \cite{Ponzano1969,cmp/1103842444}; motives and Morse theory \cite{Brown:2009ta,Bloch:2010gk}; and combinatorics of hypersphere arrangements \cite{AIF_2003__53_4_977_0,aomoto2017hypergeometric}. For reviews see \cite{Fotiadi1969,Golubeva_1976,Hwa:102287,pham1967introduction,pham2011singularities} and especially \cite{Lascoux:1968bor,Pham:1968wxy,Regge:1968rhi}.
Recent literature includes \cite{Abreu:2017ptx,Schultka:2019tfi,Mizera:2020wdt,Collins:2020euz,Muhlbauer:2020kut,Berghoff:2020bug}.

Compatibility of Landau equations with renormalization and analytic regularization has been discussed in \cite{Hepp:1966eg,Chandler:1970cy,Sato:1977nk,1977131,Muhlbauer:2020kut}. The dimension of Landau varieties was studied in \cite{Sato:1977nk}.
Their formulation in non-local theories was given in \cite{Chin:2018puw} (see also \cite{deLacroix:2018tml}), and in the flat-space limit of anti-de Sitter scattering in \cite{Maldacena:2015iua,Komatsu:2020sag}.
For a recent discussion of observable signatures of anomalous thresholds at particle colliders see \cite{Passarino:2018wix}, where we also refer the reader for a more complete list of references on the applications of Landau equations to the Standard Model physics.

\subsection{Simple Example}

\begin{figure}[!t]
	\includegraphics[scale=1.1]{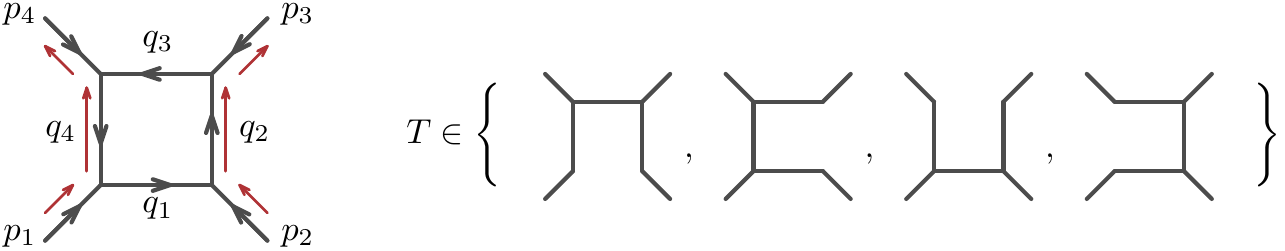}
	\caption{\label{fig:box}Planar box diagram and its spanning trees $T$. The red arrows indicate the direction of energy flow.}
\end{figure}

Let us illustrate the formulae from the previous subsections on the simple example of a box diagram with $n=4$, $\E=4$, and $\L=1$; see Fig.~\ref{fig:box}. All the internal edges and external legs are given auxiliary orientations indicated on the figure. There are four possible spanning trees $T$, one for each edge that can be clipped. According to the definition \eqref{eq:U}, this gives
\be
\U = \alpha_1 + \alpha_2 + \alpha_3 + \alpha_4.
\ee
Momentum conservation \eqref{eq:LE1} at each vertex yields
\be
p_i^\mu - q_i^\mu + q_{i-1}^\mu = 0
\ee
for $i=1,2,3,4$ with the cyclic identification $q_0^\mu := q_4^\mu$.
The continuity law \eqref{eq:LE2} around the loop reads
\be
\sum_{e=1}^{4}q_e^\mu\, \alpha_e  = 0.
\ee
Together, the above constraints form the linear Landau equations, which can be solved explicitly to give
\begin{align}\label{eq:solq12}
q_1^\mu = \frac{-p_2^\mu \alpha_2 -p_{23}^\mu\alpha_3 + p_1^\mu \alpha_4}{\alpha_1 + \alpha_2 + \alpha_3 + \alpha_4}, \qquad
&q_2^\mu = \frac{p_2^\mu \alpha_1 - p_3^\mu\alpha_3 + p_{12}^\mu \alpha_4}{\alpha_1 + \alpha_2 + \alpha_3 + \alpha_4},\\
q_3^\mu = \frac{p_{23}^\mu \alpha_1 + p_3^\mu \alpha_2 + p_{123}^\mu \alpha_4}{\alpha_1 + \alpha_2 + \alpha_3 + \alpha_4}, \qquad
&q_4^\mu = \frac{-p_1^\mu \alpha_1 - p_{12}^\mu \alpha_2 - p_{123}^\mu \alpha_3}{\alpha_1 + \alpha_2 + \alpha_3 + \alpha_4}.\label{eq:solq34}
\end{align}
Let us focus on the energy component $\mu=0$, so that the external energies $p_i^0$ have a definite sign. In the case of $12\to34$ scattering we have $p_1^0, p_2^0 >0$ and $p_3^0, p_4^0 < 0$. Together with the fact that all $\alpha_e > 0$, the above solution gives
\be
q_2^{0} > 0, \qquad q_4^{0} < 0,
\ee
while the remaining energies $q_1^0$ and $q_3^0$ do not have a definite sign. In other words, if the propagators associated to the two side edges in the diagram were ever put on-shell, the energy flowing through them is forced to propagate in the causal direction. This is in fact a general feature of planar diagrams and will be quite central to proving crossing symmetry, even though at this stage it might not be obvious why.

Let us compute the action on the support of the above solution
\be
\V(\alpha_e) = \sum_{e=1}^{4} (q_e^2 - m_e^2)\, \alpha_e \Big|_{(\ref{eq:solq12},\ref{eq:solq34})} = \frac{s\, \alpha_2 \alpha_4 + t\, \alpha_1 \alpha_3 + \sum_{i=1}^{4} \!M_i^2\, \alpha_i \alpha_{i-1} }{\alpha_1 + \alpha_2 + \alpha_3 + \alpha_4} - \sum_{e=1}^{4} m_e^2 \alpha_e
\ee
with the Mandelstam invariants $s := (p_1 {+} p_2)^2$, $t := (p_2 {+} p_3)^2$, and $\alpha_0 := \alpha_4$. It is straightforward to verify that the definition \eqref{eq:V2} together with \eqref{eq:FS} gives the same answer. The resulting system of Landau equations can be easily solved \cite{doi:10.1063/1.1703645}. Since the form of the solution itself is not hugely illuminating, let us quote it in the case where all the internal and external masses are separately equal, i.e., $m_e = m$ and $M_i = M$ (with $m \neq M/2$). It reads
\be
\big[\alpha_1^\ast : \alpha_2^\ast : \alpha_3^\ast : \alpha_4^\ast\big] = \big[ 2M^2{-}4m^2 \;:\; 4m^2{-}t \;:\; 2M^2 {-} 4m^2 \;:\; 4m^2{-}t \big]
\ee
together with
\be
st + 4m^2 u - 4M^4 = 0,
\ee
where $u := (p_1{+}p_3)^2 = 4M^2 {-} s {-} t$.
The solution can be $\alpha$-positive only when $m > M/{\sqrt{2}}$ and $t > 4m^2$ or alternatively when $m < M/{\sqrt{2}}$ and $t < 4m^2$. In either case, the solution does not intersect the physical region in the $s$- or $t$-channels, for which $st<0$ and $u<0$, but it can pass through the $u$-channel region with $u>0$ and $s,t<0$ when $m < M/2$.

One can similarly solve for the subleading Landau singularities, which we state here for completeness. For example, when $\alpha_1 = 0$ we have the triangle Landau anomalous threshold,
\be
\big[\alpha_2^\ast : \alpha_3^\ast : \alpha_4^\ast\big] = \big[ m^2 \,:\, M^2 {-} 2m^2 \,:\, m^2 \big], \qquad s\, m^2 + M^2 (M^2 {-} 4m^2) = 0,
\ee
which is $\alpha$-positive for $m < M/\sqrt{2}$ and does not intersect the $s$-channel region, but it does the other two when $m<M/2$. Similarly, when $\alpha_1 = \alpha_3 = 0$ we find two discrete solutions
\be
\big[\alpha_2^\ast : \alpha_4^\ast\big] = [1 : \pm1], \qquad s - (m\pm m)^2 = 0.
\ee
Only the $+$ solution is $\alpha$-positive and it corresponds to the normal threshold. There are analogous subleading Landau singularities for the other reduced diagrams obtained by cycling through $\alpha_i \to \alpha_{i+1}$ and $s \leftrightarrow t$. The only second-type singularity for this diagram corresponds to collinear kinematics (given by vanishing of any $3\times 3$ minor of the Gram matrix $p_i {\cdot} p_j$), yielding $s t u = 0$, which demarcates boundaries of the physical regions.

\subsection{Solution of the Linear Landau Equations}

Let us return to arbitrary Feynman diagrams. The linear Landau equations (\ref{eq:LE1},\ref{eq:LE2}) can be solved explicitly and give
\be\label{eq:qe}
q_e^\mu = \frac{1}{\U} \sum_{\substack{\text{spanning}\\ \text{trees }T}} \!\!p_{T,e}^\mu \prod_{e' \notin T} \alpha_{e'},
\ee
where $p_{T,e}^\mu$ denotes the total external momentum flowing through the edge $e$ along the spanning tree $T$ in the orientation of the edge. (In particular, if $T$ does not include $e$, we have $p_{T,e}^\mu = 0$.) In order to confirm this, let us first check that the momentum conservation \eqref{eq:LE1} at every vertex $v$ is satisfied:
\be
p_v^\mu + \sum_{e=1}^{\E} \eta_{ve} q_e^\mu = \frac{1}{\U} \sum_{T} \left(p_v^\mu + \sum_{e=1}^{\E}\eta_{ve}\, p_{T,e}^\mu \right) \prod_{e' \notin T} \alpha_{e'} = 0.
\ee
In the first equality we used the definition of $\U$ from \eqref{eq:U}. For every spanning tree $T$ the term in the parentheses vanishes because the total momentum outgoing from $v$ equals to $p_v^\mu$. In order to confirm the continuity law \eqref{eq:LE2}, let us notice that
\be
q_e^\mu \alpha_e = \frac{1}{\U}\! \sum_{T = T_1 \sqcup T_2 \sqcup e}\!\!\! p_{T,e}^\mu \prod_{e' \notin T_1, T_2} \alpha_{e'},
\ee
where the sum runs over only those spanning trees $T$ that are disjoint unions of two trees $T_1$, $T_2$, and the edge $e$ itself. Therefore the sum \eqref{eq:LE2} can be labeled by $2$-forests, which gives
\be
\sum_{e=1}^{\E} \eta_{Ie} q_e^\mu \alpha_e  = \frac{1}{\U} \sum_{T_1 \sqcup T_2} \left(\sum_{e=1}^{\E}\eta_{Ie}\, p_{T_1,T_2,e}^\mu\right)\! \prod_{e' \notin T_1, T_2} \!\! \alpha_{e'} = 0,
\ee
where $p_{T_1,T_2,e}^\mu$ denotes the total momentum flowing from one tree to the other. The sum over all edges along the loop $I$ in the parentheses vanishes because there is no net momentum flowing between $T_1$ and $T_2$ within such a loop. This concludes the proof of \eqref{eq:qe}. Note that, as expected, the result is linear in the external kinematics and at this stage all the Lorentz components are independent of each other. Using spanning trees from Fig.~\ref{fig:box} one can verify that (\ref{eq:solq12},\ref{eq:solq34}) are the solutions prescribed by \eqref{eq:qe}.

In particular, linearity in the external kinematics implies that we can treat \eqref{eq:qe} as a superposition of multiple smaller problems, where the external momenta only enter at a single source and a single sink vertex at a time. This observation will turn out to simplify the discussion in the later sections. Of course, the contributions from smaller problems get all mixed together once the quadratic Landau equations are imposed.

The above solution is essentially a theorem in the combinatorics of electrical networks; see, e.g., \cite{bollobas2013modern}. Their relevance in this problem should not come as a surprise, because both Feynman diagrams and electrical networks can be thought of as a theory of free scalar fields on a graph coupled to external sources. In the language of circuits, component-by-component $q_e^\mu$ and $\alpha_e$ are the current and resistance of the edge $e$, while the linear Landau equations (\ref{eq:LE1},\ref{eq:LE2})  are the same as the Kirchhoff's current and voltage laws respectively \cite{Bjorken:1959fd,Mathews:1959zz}. This analogy cannot be taken too far, however, because the momentum vectors we use are contracted with the Minkowski metric, so for example the quadratic Landau equations would not have a clear electric-circuit interpretation.

\subsection{\label{sec:energy-flow}Energy Flow in Planar Diagrams}

\begin{figure}[!t]
	\includegraphics[scale=1.2]{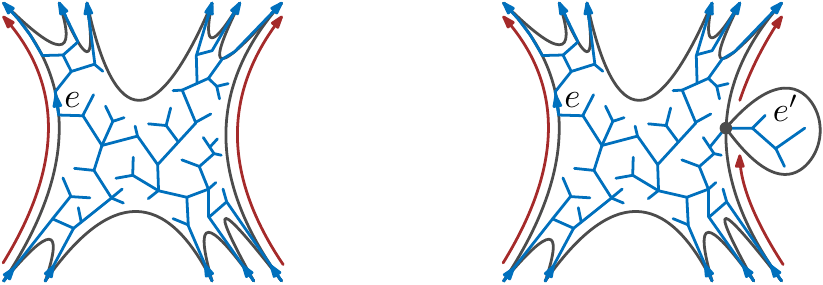}
	\caption{\label{fig:sides}Left: Diagram illustrating the flow of energy along the sides of the diagram (red), with an example spanning tree (blue) passing through one of the edges $e$ guaranteeing the causal energy flow. Right: Energies of tadpole edges $e'$ are not constrained since the net momentum flow into a tadpole is zero.}
\end{figure}

Let us turn our focus to planar Feynman diagrams. We consider situations in which the external particles are partitioned into two non-empty consecutive sets of incoming and outgoing particles, as illustrated in Fig.~\ref{fig:sides}. Given any planar embedding, let us look at the edges $e$ lying on the \emph{sides} of the diagram, i.e., the leftmost edges connecting the leftmost incoming and outgoing particles and likewise for the right side. In Fig.~\ref{fig:sides} these are indicated with the red arrows. We choose orientations of each such $e$ to be consistent with the red arrows.

Consider the energy $q_e^0$ of any such edge $e$ along the sides. According to \eqref{eq:qe}, it is given by a weighted sum over the total energies $p_{T,e}^0$ flowing through $e$ along a spanning tree $T$. For every such tree we must have $p_{T,e}^0 > 0$. This is because one cannot draw a spanning tree passing through $e$ with the opposite energy flow without violating planarity.
We conclude that $q_e^0 > 0$ for any positive values of Schwinger parameters. In other words, if there was a solution of Landau equations, it can only happen when the energies of side edges are \emph{strictly positive}. Note that for generic edges in the bulk of the diagram no such statement can be made: the energy could be positive, negative, or zero depending on the specific values of Schwinger parameters, cf. (\ref{eq:solq12},\ref{eq:solq34}).

The exceptions to the above statement are diagrams involving tadpoles, as illustrated on the right panel of Fig.~\ref{fig:sides}. Here the edges $e'$ belonging to the tadpole are not determined in terms of the external kinematics and hence must have $q_{e'}^\mu = 0$.

A physical intuition for the above result comes from the interpretation of $\V$ as the worldline action, which can be thought of as minimizing the total (Lorentzian) length of the diagram in the classical limit. Therefore a solution with energy flowing back and forth along the sides would not be optimal.

We will return to the problem of energy flow in Sec.~\ref{sec:crossing}, where it will be used to constrain the solutions of quadratic Landau equations and hence play a crucial role in the proof of crossing symmetry.

\section{\label{sec:physical}Analyticity Near the Physical Regions}

As the first step in the analytic continuation we need to discuss analyticity in the infinitesimal neighborhoods of the physical regions. Physical regions are disjoint domains of the real on-shell kinematic space labeled by the signs of energies of the external momenta, with examples illustrated in Fig.~\ref{fig:ResRet} for $n=4$. Showing analyticity near such regions will allow us to continue the scattering amplitude freely between different generic kinematic configurations within their lightcones, as indicated in the step $\RN{1}$ of Fig.~\ref{fig:steps}. The endpoint of such a deformation will be determined later. Microanalyticity of this type has the origin in macrocausality of the S-matrix \cite{Chandler:1969nd,Cutkosky:1969fq,Iagolnitzer:1977sw}, and it approximately matches the levels of analyticity proven non-perturbatively \cite{Bros:1964iho,Bros:1972jh}.

To demonstrate such analytic properties we will deform both the external kinematics and the Schwinger parameters at the same time. It will be instructive to distinguish between neighborhoods of singular and non-singular kinematic points, because they differ in which of the two deformations has the dominant effect. Planarity is not assumed in this section.

\subsection{\label{sec:non-singular}Neighborhoods of Non-Singular Points}

We start with the neighborhoods of non-singular points, i.e., those for which Landau equations do not have solutions. However, above production thresholds it can still happen that $\V = 0$ somewhere along the integration contour. Such a singularity can be resolved in many ways by a deformation of the contour, resulting in different values for the integral. The $i\eps$ prescription reminds us which contour deformation should be taken to compute Feynman integrals associated with causal scattering processes.

Contour deformation can be implemented by giving small phases to the Schwinger parameters, amounting to a change of variables $\alpha_e \mapsto \check{\alpha}_e$. There exists a canonical prescription for such a deformation:
\be
\check{\alpha}_e := \alpha_e\, e^{i \eps (q_e^2 - m_e^2)},
\ee
where, as before, $q_e^\mu = q_e^\mu(\alpha_{e'},p_i)$ are the solutions of the linear Landau equations and hence functions of the Schwinger parameters and the external kinematics as given in \eqref{eq:qe}. Here $\eps$ is a small parameter. The endpoints of integration are preserved and the Jacobian for this transformation is non-singular. As a result, the deformed action $\check{\V} = \V(\check{\alpha}_e, p_i)$ becomes
\be
\check{\V} = \sum_{e=1}^{\E} (\check{q}_e^2 - m_e^2)\, \check{\alpha}_e,
\ee
where $\check{q}_e^\mu = q_e^\mu(\check{\alpha}_{e'},p_i)$ are evaluated on the rotated  $\check{\alpha}_{e'}$ instead of $\alpha_{e'}$. Since to leading orders the Schwinger parameters have the expansion
\be
\check{\alpha}_e = \alpha_e + i\eps (q_e^2 - m_e^2)\, \alpha_e + {\cal O}(\eps^2),
\ee
the deformed action $\check{\V}$ can be Taylor expanded to give
\begin{align}
\check{\V} &= \V + i\eps \sum_{e=1}^{\E} (q_e^2 - m_e^2)\, \alpha_e\, \frac{\partial \V}{\partial \alpha_e}  + {\cal O}(\eps^2)\nn\\
&= \V + i\eps \sum_{e=1}^{\E} (q_e^2 - m_e^2)^2\, \alpha_e + {\cal O}(\eps^2).\label{eq:checkV}
\end{align}
When the kinematics is real, the coefficient of $i\eps$ is always non-negative. Moreover, it is zero if and only if the quadratic Landau equations $q_e^2 = m_e^2$ are satisfied for all $e$, which does not happen by assumption. We conclude that for sufficiently small $\eps$ the above contour deformation implements the correct $i\eps$ prescription. In fact, this is the principle underlying modern numerical approaches to the evaluation of Feynman integrals, see, e.g., \cite{Nagy:2006xy,Anastasiou:2007qb}, though its simplicity is often obscured by unfortunate choices of the $\mathrm{GL}(1)$ gauge fixing.

To complete the arguments we need to show that there exists a complex neighborhood of the kinematic point where the above $i\eps$ prescription still works. This is clearly the case if we deform the external kinematics $p_i^\mu \mapsto \hat{p}_i^\mu$ with some complex displacement according to
\be\label{eq:phat}
\hat{p}_i^\mu := p_i^\mu + \eps^{2} \Delta p_i^\mu,
\ee
such that it still preserves momentum conservation and the mass-shell conditions. This shift is chosen to be subleading to the contour deformation. The internal momenta react to this change according to \eqref{eq:qe} and hence can be written as 
\be\label{eq:qhat}
\hat{q}_e^\mu = q_e^\mu + \eps^2 \Delta q_e^\mu
\ee
prior to the contour deformation. After contour deformations this change affects $\check{\V}$ only at the subleading order ${\cal O}(\eps^2)$. It implies that for sufficiently small $\eps$ there exists a complex neighborhood of any non-singular point where the amplitude is analytic. Note that we do not attempt to optimize for the size of the analyticity region, since in our applications it suffices that an infinitesimal neighborhood exists. Concrete bounds on the reminder coefficients were given in \cite{Chandler:1969nd}.

\begin{figure}[!t]
	\includegraphics[scale=1.2]{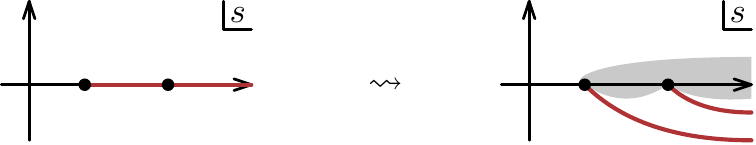}
	\caption{\label{fig:physical}Schematic diagram illustrating the domain of analyticity near a physical region (for example, in the center of mass energy variable $s$) for a planar process. Black dots and red curves denote branch points and cuts respectively. Physical region is a subset of the real axis. After using the $i\eps$ prescription, branch cuts have been deformed by an infinitesimal amount. The domain of analyticity is denoted by the shaded region and includes complex neighborhoods of non-singular points, as well as those of the singular ones, but only from a specific direction prescribed by \eqref{eq:ieps-rule}. This analysis does not constrain singularities in the unshaded regions.}
\end{figure}

The result of this procedure is schematically illustrated in Fig.~\ref{fig:physical}, where the $i\eps$ deformation moves branch cuts by a small amount and the amplitude is analytic in a neighborhood of non-singular points parametrically smaller than the branch cut deformation.
(Recall that positions of branch cuts are physically meaningless: they only reflect the fact that we cannot represent the kinematic space for complexified scattering amplitudes, which are in general multi-valued functions, on a piece of paper.) Out of different ways of resolving the $\V=0$ ambiguity, the causal $i\eps$ prescription is the one that guarantees we end up with the scattering amplitude on the correct sheet.

\subsection{Neighborhoods of Singular Points}

The above discussion guarantees that there are no singularities in the neighborhood of the connected components of the physical regions (cf. gray areas in Fig.~\ref{fig:ResRet}), however it still does not explain how to continue around singularities (cf. black curves crossed by the path $\RN{1}$ in Fig.~\ref{fig:ResRet}). Singular points turn out to have neighborhoods of analyticity, but only when approached from a specific side and hence require a more careful treatment.

Solutions of Landau equations with $\alpha_e>0$ generically lie on codimension-one curves in the physical regions; see, e.g., \cite{Iagolnitzer:1994xv}. Although there might exist points on intersections of multiple such singularities (once all the subleading Landau curves are included too), they are of higher codimension and hence can be always avoided when the space-time dimension is $\D > 2$. (At such special points, scattering amplitudes cannot, in principle, be represented as boundary values of a single analytic function; see \cite{Chandler:1969nd} for an explicit example and \cite{Iagolnitzer:1977sw} for a connection to no particle production in two-dimensional theories.)

Therefore, without loss of generality it will be sufficient to study what happens when we cross a single codimension-one curve. In the Schwinger parameter space it is determined by the solution at some specific point $\alpha_e^\ast$, or rather the equivalence class of such points up to $\alpha_e^\ast \sim \lambda \alpha_e^\ast$. Let us denote the internal momenta $q_e^\mu$ evaluated at this point with $(q_e^\ast)^\mu = q_e^\mu(\alpha_{e'}^\ast,p_i^\ast)$, where $(p_i^\mu)^\ast$ is the singular kinematics. By definition, they solve Landau equations and hence satisfy
\be\label{eq:on-shell}
(q_e^\ast)^2 = m_e^2
\ee
for all $e$ simultaneously, which also means that $\V^\ast = \V(\alpha_e^\ast, p_i^\ast) = 0$. As explained in the previous subsection, we cannot perform any contour deformations to escape this singularity because $\check{\alpha}_e^\ast = \alpha_e^\ast$ and as a consequence $\check{\V}^\ast = \V(\check{\alpha}_e^\ast, p_i^\ast) =0$ identically.

We can however deform the external kinematics according to \eqref{eq:phat}. The response of the Schwinger parameters to simultaneous contour and kinematic deformation is
\be
\hat{\check{\alpha}}_e^\ast = \alpha_e^\ast\, e^{i\eps ((\hat{q}_e^\ast)^2 - m_e^2)} = \alpha_e^\ast + {\cal O}(\eps^3),
\ee
where we used \eqref{eq:qhat} and \eqref{eq:on-shell}.
Therefore the action $\hat{\check{\V}}^\ast = \V(\check{\alpha}_e^\ast, \hat{p}_i^\ast)$ expanded around such a point gives to leading orders
\begin{align}\nn
\hat{\check{\V}}^\ast &= \sum_{e=1}^{\E} ((\hat{\check{q}}_e^\ast)^2 - m_e^2)\, \hat{\check{\alpha}}_e^\ast \\
&= 2\eps^{2} \sum_{e=1}^{\E} \Delta q_e^\ast {\cdot} q_e^\ast\, \alpha_e^\ast + {\cal O}(\eps^3),
\end{align}
where all the contributions from contour deformations are subleading and only the kinematic ones matter at the leading order $\eps^2$.
Since all the integration variables are localized, the leading factor is purely a function of the external kinematics, which vanishes on the Landau curve. In fact, it can be thought of as an implicit parameterization for such a singularity. It is now clear that the neighborhood of this curve that is free of singularities and consistent with the $i\eps$ prescription has to be given by
\be\label{eq:ieps-rule}
\Im \left(\sum_{e=1}^{\E} \Delta q_e^\ast {\cdot} q_e^\ast\, \alpha_e^\ast \right) > 0
\ee
for sufficiently small $\eps$.

For example, in terms of the Mandelstam variables for $n=4$ the deformation implies a shift $\hat{s} = s^\ast + \eps^2 \Delta s + {\cal O}(\eps^4)$ and $\hat{t} = t^\ast + \eps^2 \Delta t + {\cal O}(\eps^4)$, where $(s^\ast, t^\ast)$ denotes the singular point. Using \eqref{eq:V2}, around this singularity for planar anomalous and normal thresholds we have
\be
\hat{\check{\V}}^\ast = \eps^2 \left( \Delta s\, \F_{12}^\ast  + \Delta t\, \F_{23}^\ast \right) + {\cal O}(\eps^3),
\ee
which is always causal in the directions containing $\Im \Delta s > 0$ and $\Im \Delta t > 0$ with sufficiently small $\eps$, since $\F_S$ are always positive according to \eqref{eq:FS}.

We refer interested readers to the literature on positive-$\alpha$ Landau singularities, where microanalyticity of this type has been studied in greater technical detail in a somewhat similar language \cite{AIHPA_1967__6_2_89_0,Iagolnitzer:1968zz,doi:10.1063/1.1664486,Iagolnitzer1969,Chandler:1969bd,PhysRev.174.1749,Iagolnitzer:1994xv}.

\section{\label{sec:crossing}Analyticity in the Crossing Domains}

In this section we consider analytic continuation between the future and past lightcones. They will be connected through specific regions of the complexified kinematic space we call \emph{crossing domains}. We will first identify all the potential singularities that could contribute along such a deformation and then show their existence leads to a contradiction with the energy flow in planar Feynman diagrams. This result will allow us to complete the deformation indicated in Fig.~\ref{fig:steps}.

Before considering crossing domains generally, let us focus on the simplest case $n=4$, which already illustrates all the essential features of the derivation while avoiding unnecessary proliferation of variables. We will be working in the lightcone coordinates $p^\mu = (p^+, p^-, \vec{p})$ with the Lorentz norm $p^2 = p^+ p^- - \vec{p}^{\,2}$. Planarity is not assumed until the very end.

\subsection{Four-Point Example}

Let us first work with an arbitrary $n=4$ Feynman diagram. In order to define the starting point of the step $\RN{1}$, we commit to the Lorentz frame in which the external momentum vectors are written as
\begin{gather}\label{eq:4pt}
p_1^\mu = \big(p_1^+,\, p_1^-,\, \vec{p}_1\big),\quad\;
p_2^\mu = \big(p_2^+,\, p_2^-,\, \vec{p}_2\big),\\
p_3^\mu = \big({-}p_2^+,\, {-}p_2^-,\, \vec{p}_3\big),\quad\;
p_4^\mu = \big({-}p_1^+,\, {-}p_1^-,\, \vec{p}_4\big),
\end{gather}
where the components orthogonal to the lightcone satisfy momentum conservation $\sum_{i=1}^{4} \vec{p}_i = 0$ and we require $\vec{p}_2 \neq -\vec{p}_3$ so that the kinematics is not collinear. The mass-shell conditions read $M_i^2 = p_i^+ p_i^- - \vec{p}_i^{\,2}$. In this frame, the lightcone components of the particles $2$ and $3$ are diametrically opposed, and likewise for $1$ and $4$.
We choose the energies to satisfy
\be
p_1^\pm > 0, \qquad p_2^\pm > 0,
\ee
which corresponds to $12 \to 34$ scattering. In particular, for massless particles we require that they are both non-zero.
Such a frame always exists and even leaves a possibility for a further boost (e.g. to $p_1^+ = p_1^-$) that we will not use. As before, we require that such kinematics is non-singular. It defines the beginning of the step $\RN{1}$.

In order to describe the endpoint of step $\RN{1}$ and the beginning of step $\RN{2}$, we will simply require that the momenta of particles $2$ and $3$ lie closer to the positive axis of the lightcone than the remaining vectors. In other words,
\be\label{eq:closer}
\frac{p_1^+}{p_1^-} < \frac{p_2^+}{p_2^-}.
\ee
We can clearly make small wiggles around such kinematic points that preserve the constraint, so the whole discussion can be repeated in an open set of the path we will describe, which is necessary for analytic continuation. Note that all the Mandelstam invariants remain finite. Of course, it might be that the original kinematics already satisfies \eqref{eq:closer}. If this is not the case, the scattering amplitude can be analytically continued via path $\RN{1}$ in the neighborhood of the physical region, as described in Sec.~\ref{sec:physical}. Note that this step requires $\D > 2$.

\subsubsection{Rotation in the Complexified Lightcones}

\begin{figure}[!t]
	\includegraphics[scale=1.2]{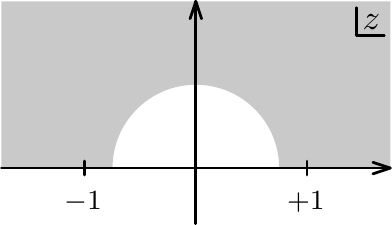}
	\caption{\label{fig:z} Region of analyticity in the $z$-plane (shaded). Analytic continuation between $z=1$ and $z=-1$ in the upper-half plane corresponds to flipping the signs of energies of selected particles.}
\end{figure}

The kinematic deformation performed in step $\RN{2}$ is simply a rotation to the other side of the lightcone for particles $2$ and $3$. Denoting the deformed variables with hats, we take
\be\label{eq:deformation-n4}
\hat{p}_2^\mu = \big(z p_2^+,\, \mfrac{1}{z} p_2^-,\, \vec{p}_2\big),\qquad
\hat{p}_3^\mu = \big({-}z p_2^+,\, {-}\mfrac{1}{z} p_2^-,\, \vec{p}_3\big)
\ee
with the remaining components untouched. This deformation preserves momentum conservation and mass-shell conditions. We will use the simplest possible deformation between $z=1$ and $z=-1$ along a path in the upper-half plane, $\Im z > 0$, outside of the unit semi-circle, $|z|^2 \geq 1$, as illustrated in Fig.~\ref{fig:z}. In terms of the Mandelstam invariants this translates to
\begin{align}
\Im \hat{s} &= \Im\, (p_1 {+} \hat{p}_2)^2 \nn\\
&= \Im z \left( p_2^+ p_1^- - \mfrac{1}{\,|z|^2}\, p_1^+ p_2^- \right) > 0,\label{eq:Ims}
\end{align}
where we used the condition \eqref{eq:closer} for the inequality, and similarly
\be
\Im \hat{t} = \Im \, (\hat{p}_2 {+} \hat{p}_3)^2 = 0,
\ee
since $\hat{t} = t$ remains undeformed. Because all the masses are real, the scattering process in the \emph{imaginary} directions looks like a highly-energetic process with vanishing squared momentum transfer and masses, even though in reality all the energies involved are finite \footnote{Note that we could have done away without the constraint \eqref{eq:closer} by deforming into $\Im z < 0$ and $|z|^2 \leq 1$ when it is not satisfied. However, aside from complicating the discussion, there is no simple generalization to higher multiplicity.}.

\subsubsection{Absence of Singularities}

In order to see how this deformation acts on the internal momenta $\hat{q}_e^\mu = q_e^\mu(\alpha_{e'},\hat{p}_i)$, let us first use the linear Landau equations. It will not be necessary to deform Schwinger parameters. Only the components $\mu = \pm$ are affected. Due to linearity, we can separately consider the effect of the momenta of the particles $1$ and $4$ from that of the particles $2$ and $3$. Therefore, the internal momenta on the solutions to linear Landau equations can be written as
\be
\hat{q}_e^{\pm} = p_1^{\pm} f_{e,14} + z^{\pm 1} p_2^{\pm} f_{e,23}.
\ee
Here we have stripped the kinematic dependence from the factors $f_{e,ij}$. They can be expressed in terms of sums over spanning trees $T$, as given in \eqref{eq:qe},
\be\label{eq:feij}
f_{e,ij} := \frac{1}{\U} \sum_T \eta_{e,ij}^T \prod_{e' \notin T} \alpha_{e'},
\ee
where $\eta_{e,ij}^T$ equals to $+1$ ($-1$) if the unique path from $i$ to $j$ along the spanning tree $T$ passes through $e$ with the same (opposing) orientation and $0$ otherwise.

Using this solution, the imaginary parts of the quadratic Landau equations read
\be\label{eq:qLE}
\Im (\hat{q}_e^2 - m_e^2) = \Im z \left( p_2^+ p_1^- - \mfrac{1}{\,|z|^2}\, p_1^+ p_2^- \right) f_{e,14}\, f_{e,23} = 0.
\ee
The fact that they are proportional to $\Im \hat{s}$ should not be surprising because it is the only mass scale in the problem. In order to prove analyticity in the region indicated in Fig.~\ref{fig:z}, we need to show that \eqref{eq:qLE} cannot have solutions for all $e$ simultaneously. Let us first understand how such a solution would have to look like \emph{if} it existed.

\begin{figure}[!t]
	\includegraphics[scale=1.1]{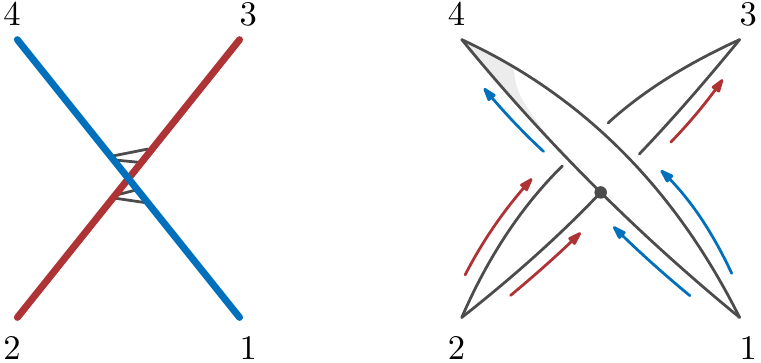}
	\caption{\label{fig:beams}Left: Singularity associated to a potential solution of Landau equations of the type \eqref{eq:aligned} corresponding to two beams of particles aligned with $p_1^\pm$ (blue) and $p_2^\pm$ (red) scattering off each other. Right: One-vertex reducible diagram, for which the imaginary parts of Landau equations are trivially satisfied, but still cannot be singular.}
\end{figure}

The constraints \eqref{eq:qLE} require that in the imaginary directions each propagator looks like a massless particle placed on-shell.
It can be achieved when either $f_{e,14}=0$, or $f_{e,23}=0$, or when both are true at the same time. They respectively correspond to the solutions for which
\be\label{eq:aligned}
\hat{q}_e^\pm \,\propto\, p_2^{\pm} \qquad\mathrm{or}\qquad \hat{q}_e^\pm \,\propto\, p_1^{\pm} \qquad\mathrm{or}\qquad \hat{q}_e^\pm \,=\, 0,
\ee
that is, each internal momentum $\hat{q}_{e}^\pm$ in the lightcone directions has to be aligned with either of the external momenta or vanish. Moreover, by momentum conservation the edges belonging to each category have to be connected to each other and to the respective external legs. In the last case, $\hat{q}_e^\pm = 0$, the real on-shell condition reads $-\vec{q}_e^{\,2}-m_e^2 = 0$, which can only ever be attained when $m_e = 0$. Therefore if a singularity was to develop, it would have to look like two beams of particles, aligned in the $2$--$3$ and $1$--$4$ directions, scattering off each other and possibly exchanging massless states, see the left panel of Fig.~\ref{fig:beams}.

As a consequence, analyticity in the crossing domains is contingent upon proving that singularities of the type explained above are absent. Note that thus far we have not made any assumptions on planarity. It is only used in the following final step.

\begin{figure}[!t]
	\includegraphics[scale=1.1]{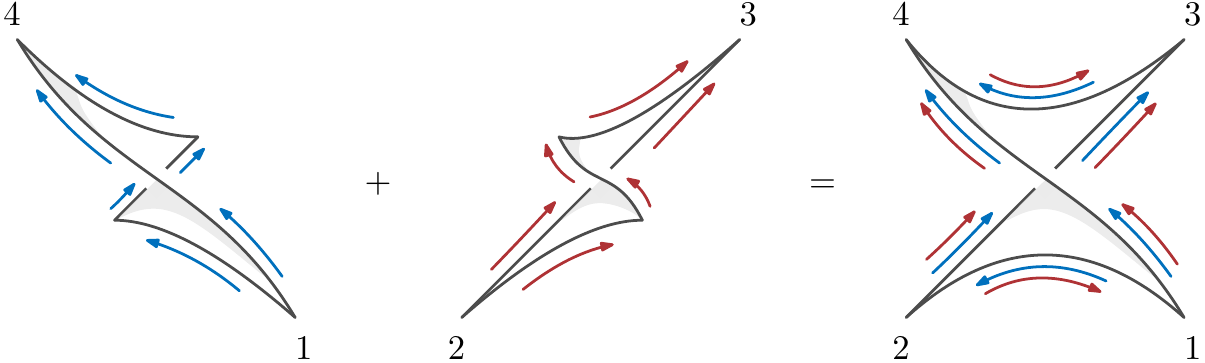}
	\caption{\label{fig:flow}Flow of the momenta in the directions of $p_1^\pm$ (blue) and $p_2^\pm$ (red). The edges on the sides of the diagram always have to have non-zero components in both directions, thus preventing formation of singularities in the crossing domain.}
\end{figure}

Given what we have learned in Sec.~\ref{sec:energy-flow}, the singularity from the left panel of Fig.~\ref{fig:beams} can never happen for planar Feynman diagrams. Let us pick any planar embedding of the diagram. The discussion of Sec.~\ref{sec:energy-flow} still applies to $f_{e,14}$ and $f_{e,23}$ because the diagram remains planar, but now we have two problems sourced by one incoming and one outgoing momentum each. In the above manipulations, $f_{e,14}$ is proportional to the energy flowing through this diagram as if it was sourced only by the momenta of particles $1$ and $4$. Therefore, along the side edges the components of $\hat{q}_e^\pm$ proportional to $p_1^\pm$ have to flow in a definite direction according to the blue arrows on Fig.~\ref{fig:flow}. Likewise, $f_{e,23}$ measures the energy flowing through the diagram as if it was sourced only by the momenta of the particles $2$ and $3$. For the side edges the components of $\hat{q}_e^\pm$ along the $p_2^\pm$ direction must have a definite sign according to the red arrows in Fig.~\ref{fig:flow}.

This leads us to conclude that such edges $e$ cannot be simultaneously aligned with the outgoing particles and on-shell, i.e.,
\be
\Im(\hat{q}_e^2 - m_e^2) \neq 0,
\ee
because $f_{e,14}f_{e,23}$ has a definite sign along the integration contour with positive values of Schwinger parameters. This proves analyticity in the region indicated in Fig.~\ref{fig:z} for planar diagrams.

The only exception to the above arguments are one-vertex reducible diagrams, such that removal of a single vertex disconnects the external particles $1$ and $4$ from $2$ and $3$. These are precisely the cases corresponding to the tadpoles from Fig.~\ref{fig:sides}, where Landau equations on the one side of the diagram decouple from those on the other, allowing for a possible solution of the type illustrated on the right panel of Fig.~\ref{fig:beams}. They are special cases in which the two beams of particles interact solely at a single point. However, such diagrams only depend on the squared momentum transfer $t$ and are $s$-independent. Since along the path of deformation $t$ remained fixed, it means that if there were no singularities at the starting point of the deformation, there are no singularities in the crossing domain either.

\subsubsection{Compatibility with the Physical Regions}

In order to complete the discussion, we need to show compatibility with Sec.~\ref{sec:physical}, i.e., that the physical regions are approached from the correct direction at $z=1$ and $z=-1$. (Strictly speaking, this is not required near $z=-1$, but turns out to be true in planar cases anyway). This is rather cumbersome to see from the loop momentum picture, so we turn back to analyzing the action $\V$. To this end we use the representation \eqref{eq:V2}, which for planar diagrams at $n=4$ reads
\be
\V = s\, \F_{12} + t\, \F_{23} + \sum_{i=1}^{4} M_i^2 \F_{i} - \sum_{e=1}^{4} m_e^2 \alpha_e.
\ee
Under the deformation \eqref{eq:deformation-n4} only the Mandelstam invariant $s$ acquires an imaginary part and hence we have
\be\label{eq:ImV}
\Im \hat{\V} = \Im \hat{s}\, \F_{12} > 0,
\ee
where we used the fact that $\Im \hat{s} > 0$ in the region indicated in Fig.~\ref{fig:z} according to \eqref{eq:Ims} and likewise $\F_{12} > 0$ from its definition in \eqref{eq:FS}. Therefore approaching $z=1$ and $z=-1$ from the upper-half $z$-plane defines the correct causal prescription. Note that \eqref{eq:ImV} by itself guarantees analyticity in the shaded region in Fig.~\ref{fig:z}, however it does not give the same physical intuition that the analysis in terms of Lorentz vectors gave us. Even more so, it is straightforward to see that planar diagrams are always analytic when $\Im s >0$ and/or $\Im t >0$. Finally, note that the $u$-channel physical region is approached from the correct $i\eps$ direction, which in general is not the case for non-planar diagrams.

Once the step $\RN{2}$ of the analytic continuation has been completed, the remaining ones can be performed using the same methods with permuted labels. Reading the Fig.~\ref{fig:steps} backwards, steps $\RN{5}$ and $\RN{4}$ are the same as $\RN{1}$ and $\RN{2}$ up to the relabeling $(1234) \to (2\bar{3}4\bar{1})$ and $s \leftrightarrow t$. The step $\RN{3}$ exchanges the positions of $1$ and $\bar{3}$ in the neighborhood of the $u$-channel physical region using the $i\eps$ prescriptions described in Sec.~\ref{sec:physical}. Composition of the five steps gives the path of analytic continuation showing crossing symmetry for $n=4$ planar scattering amplitudes in perturbation theory.

\subsection{\label{sec:definition}Definition of Crossing Domains}

The generalization to arbitrary multiplicity $n$ is rather straightforward. For any Feynman diagram we group the incoming particles into non-empty sets $\AA$ and $\BB$, and similarly the outgoing ones into non-empty sets $\CC$ and $\DD$. Denoting with $p_S^\mu$ the total momentum of the particles in the set $S$, we pick a Lorentz frame where
\begin{gather}
p_\AA^\mu = \big(p_\AA^+,\, p_\AA^-,\, \vec{p}_\AA\big), \quad\;
p_\BB^\mu = \big(p_\BB^+,\, p_\BB^-,\, \vec{p}_\BB\big), \\
p_\CC^\mu = \big({-}p_\BB^+,\, {-}p_\BB^-,\, \vec{p}_\CC\big), \quad\;
p_\DD^\mu = \big({-}p_\AA^+,\, {-}p_\AA^-,\, \vec{p}_\DD\big),
\end{gather}
with $\vec{p}_\BB \neq - \vec{p}_\CC$. In particular, the lightcone components satisfy
\be
p_\AA^\pm := \sum_{a \in \AA} p_a^\pm = -\sum_{d \in \DD}p_d^\pm > 0
\ee
and
\be
p_\BB^\pm := \sum_{b \in \BB} p_b^\pm = -\sum_{c \in \CC}p_c^\pm > 0
\ee
with all $p_a^\pm, p_b^\pm >0$ and $p_c^\pm, p_d^\pm <0$. From now on the indices $a,b,\ldots$ will always refer to the particles in the sets $\AA, \BB, \ldots$ specifically. The above kinematics defines the starting point of step $\RN{1}$ and we assume that the scattering amplitude exists in an open set including such a point.

\begin{figure}[!t]
	\includegraphics[scale=1.2]{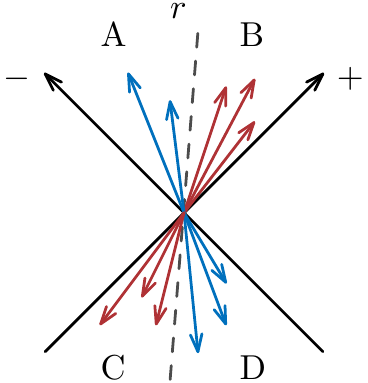}
	\caption{\label{fig:lightcone}Configuration of external momentum vectors $p_i^\mu$ in the lightcone at the end of step $\RN{1}$, with a reference line $r$ separating the four sets of particles.}
\end{figure}

In order to state the endpoint of step $\RN{1}$, let us introduce the ratio
\be
\theta_S := \frac{p_S^+}{p_S^-}
\ee
for any set of particles $S$. It is essentially the exponential of the rapidity of $p_S^\mu$, measuring how close such a momentum is to the positive axis of the lightcone. We are going to introduce a reference line passing through the origin of the lightcone, as in Fig.~\ref{fig:lightcone}, and require that the individual momenta lie on either side of it according to
\be\label{eq:ratios}
\theta_d,\, \theta_a\, <\, \theta_b,\, \theta_c.
\ee
We do not require that the individual momenta are ordered within each set.
The reference line itself will not play any role below. Continuation from an arbitrary configuration of the momenta in the lightcone and the one organized according to \eqref{eq:ratios} can be achieved using the $i\eps$ deformation explained in Sec.~\ref{sec:physical}. Since \eqref{eq:ratios} is an open condition, there exists an open set around the path of deformation we will describe.

We will continue the particles from the set $\BB$ from the future to the past lightcone, and likewise those from $\CC$ from the past to the future lightcone. To this end we deform all the momenta belonging to the two sets with a complex parameter $z$ according to
\be\label{eq:deformation}
\hat{p}_b^\mu = \big(z p_b^+,\, \mfrac{1}{z} p_b^-,\, \vec{p}_b\big), \qquad \hat{p}_c^\mu = \big({-}z p_c^+,\, {-}\mfrac{1}{z} p_c^-,\, \vec{p}_c\big)
\ee
for every $b\in \BB$ and $c\in \CC$. The deformation preserves momentum conservation, on-shell conditions, and stays within the same frame we chose above. The analytic continuation between $z=1$ and $z=-1$ will take place through the crossing domain, with $\Im z >0$ and $|z|^2 \geq 1$, as illustrated in Fig.~\ref{fig:z}. 

\subsection{Worldline Perspective}

Let us first understand analyticity in the crossing domains from the perspective of the worldline action. We consider a planar process with the cyclic ordering $(\AA\BB\CC\DD)$ such that the incoming and outgoing particles are consecutive. As the first step, we need to discuss which of the planar Mandelstam invariants $p_S^2$ are deformed and which are not. (As mentioned before, for $n > \D{+}1$ they are not independent of each other.) In order to avoid double-counting, we can take $S$ to always contain the first particle from the set $\BB$. A given $S$ consists of the particles whose momenta are not deformed under \eqref{eq:deformation} and those that are, respectively:
\be
S_{\DD\AA} := S \cap (\DD \cup \AA), \qquad S_{\BB\CC} := S \cap (\BB \cup \CC),
\ee
such that $S = S_{\DD\AA} \cup S_{\BB\CC}$. The imaginary parts of $\hat{p}_S^2$ can therefore be written as
\be
\Im \hat{p}_S^2 = \Im z \left( p_{S_{\BB\CC}}^+ p_{S_{\DD\AA}}^- - \mfrac{1}{\,|z|^2}\, p_{S_{\DD\AA}}^+ p_{S_{\BB\CC}}^- \right).
\ee
Therefore only the Mandelstam invariants with $S_{\BB\CC} \notin \{\varnothing, \BB\cup \CC\}$ and $S_{\DD\AA} \notin \{\varnothing, \DD\cup \AA\}$ are deformed. Moreover, because of planarity, if any particle from the set $\DD$ is included, so must be all those from the set $\AA$, and likewise if any particle from the set $\CC$ is included, so must be all those from the set $\BB$. We only consider such $S$ from now on.

\begin{figure}[!t]
	\includegraphics[scale=1]{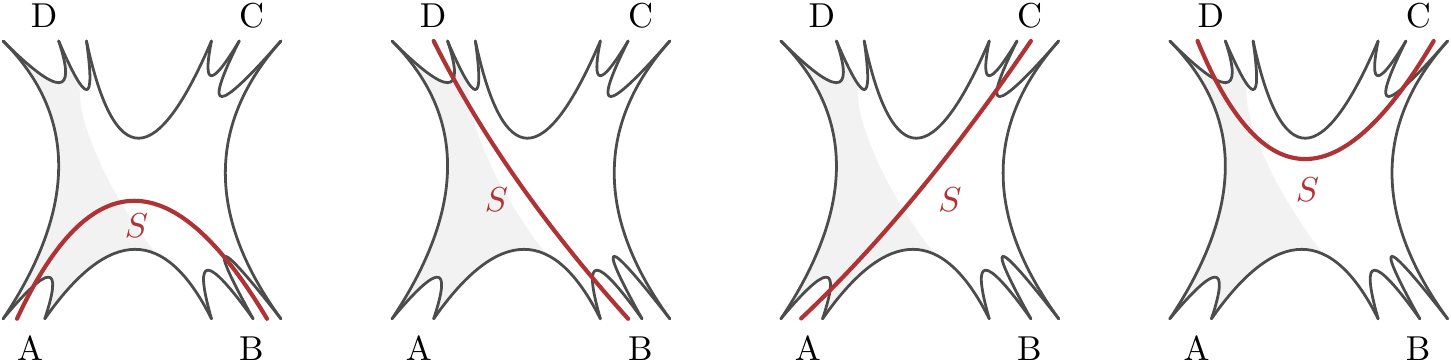}
	\caption{\label{fig:invariants}Deformed Mandelstam invariants fall into four classes illustrated above. In each case the set $S$ is the one lying below the red line and always contains at least one label from $\AA$ and $\BB$ each.}
\end{figure}

This leaves us with four possibilities, depending on whether $S$ has non-zero overlap with $\CC$ and with $\DD$, as summarized in Fig.~\ref{fig:invariants}. In the first case
\be
S\cap \CC = \varnothing,\quad S\cap \DD = \varnothing: \qquad \Im \hat{p}_S^2 = \Im z \sum_{\substack{a \in S \cap \AA\\ b \in S\cap \BB}} p_a^- p_b^- \left( \theta_b - \mfrac{1}{\,|z|^2}\, \theta_a \right) > 0,
\ee
where the constraints \eqref{eq:ratios} guarantee that each term in the sum is positive when $|z|^2 \geq 1$, which together with $\Im z > 0$ guarantees positivity of the whole expression. Similarly, the second case is
\be
S\cap \CC = \varnothing,\quad S\cap \DD \neq \varnothing: \qquad \Im \hat{p}_S^2 = -\Im z \!\!\! \sum_{\substack{d \in \DD \setminus (S \cap \DD)\\ b \in S\cap \BB}} \!\!\! p_b^- p_d^- \left( \theta_b - \mfrac{1}{\,|z|^2}\, \theta_d \right) > 0.
\ee
In the first equality we used momentum conservation to write $p_{S_{\DD\AA}}^\pm = -p_{\DD\setminus (S\cap \DD)}^\pm$. Positivity of each term in the parentheses follows from \eqref{eq:ratios}. The sum is also positive after recalling that $p_b^- p_d^- < 0$. Using exactly the same steps we have
\be
S\cap \CC \neq \varnothing,\quad S\cap \DD = \varnothing: \qquad \Im \hat{p}_S^2 = -\Im z \!\!\! \sum_{\substack{a \in S \cap \AA\\ c \in \CC \setminus (S \cap \CC)}} \!\!\! p_a^- p_c^- \left( \theta_c - \mfrac{1}{\,|z|^2}\, \theta_a \right) > 0,
\ee
using the fact that $p_{S_{\BB\CC}}^\pm = -p_{\CC\setminus(S\cap \CC)}^\pm$ and $p_c^- p_d^- < 0$ together with \eqref{eq:ratios}.
Finally, in the fourth case we apply momentum conservation twice to find
\be
S\cap \CC \neq \varnothing,\quad S\cap \DD \neq \varnothing: \qquad \Im \hat{p}_S^2 = \Im z \!\!\! \sum_{\substack{d \in \DD \setminus (S \cap \DD)\\ c \in \CC \setminus (S \cap \CC)}} \!\!\! p_c^- p_d^- \left( \theta_c - \mfrac{1}{\,|z|^2}\, \theta_d \right) > 0.
\ee
Therefore all the Mandelstam invariants are deformed in such a way that their imaginary parts remain positive in the shaded region from Fig.~\ref{fig:z}. For instance, when $n=5$ and $\AA=1$, $\BB=2$, $\CC=34$, $\DD=5$, the Mandelstam invariants $\hat{p}_{12}^2$ and $\hat{p}_{123}^2$ are deformed in the upper-half planes, while $p_{23}^2$, $p_{512}^2$, and $p_{234}^2$ remain constant.

It is now a matter of plugging this result into the worldline action $\V$ from \eqref{eq:V2} to obtain
\be\label{eq:ImV2}
\Im \hat{\V} = \sum_{S} \Im \hat{p}_S^2\, \F_S > 0,
\ee
since all the $\F_S$ are positive according to the definition \eqref{eq:FS} with positive Schwinger parameters and the sum runs over only the deformed planar Mandelstam invariants. This fact at the same time proves analyticity in the region $\Im z > 0$ and $|z|^2 \geq 1$, as well as that the physical regions are approached from the causal direction as $\Im z \to 0^+$.

We are left with the edge case that all the factors $\F_S$ appearing in \eqref{eq:ImV2} vanish identically. In such situations the amplitude would be entirely independent of the deformation parameter $z$ and hence analytic due to the initial assumption that it exists at $z=1$. This completes the proof of crossing symmetry for planar scattering amplitudes in perturbation theory between the channels with consecutive sets of incoming and outgoing particles.

\subsection{Loop Momentum Perspective}

Note that in the above proof it was not even necessary to write down the individual Landau equations. In order to gain some physical intuition behind this result, let us identify how the potential singularities in the crossing domains have to look like in terms of the momentum vectors and explain why they do not appear for planar diagrams.

We begin with an arbitrary Feynman diagram without restrictions on planarity. Making use of the superposition property of the linear Landau equations in the external kinematics we can write
\be\label{eq:qepm}
\hat{q}_e^{\pm} = -\frac{1}{p_\AA^{\pm}}\sum_{\substack{a \in \AA\\ d \in \DD}} p_{a}^{\pm} p_{d}^{\pm} f_{e,ad} - \frac{z^{\pm1}}{p_\BB^\pm} \sum_{\substack{b \in \BB\\ c \in \CC}} p_{b}^{\pm}  p_{c}^{\pm} f_{e,bc},
\ee
where $f_{e,ad}$ measure the flow of the lightcone momenta as if the diagram was only probed by a unit momentum flowing from the incoming vertex $a$ and the outgoing one $d$, and similarly for $f_{e,bc}$. The explicit expression was given in \eqref{eq:feij}. These factors only depend on the topology of the diagram, while all the kinematic information has been explicitly factored out as coefficients in \eqref{eq:qepm}.

\begin{figure}[!t]
	\includegraphics[scale=1.1]{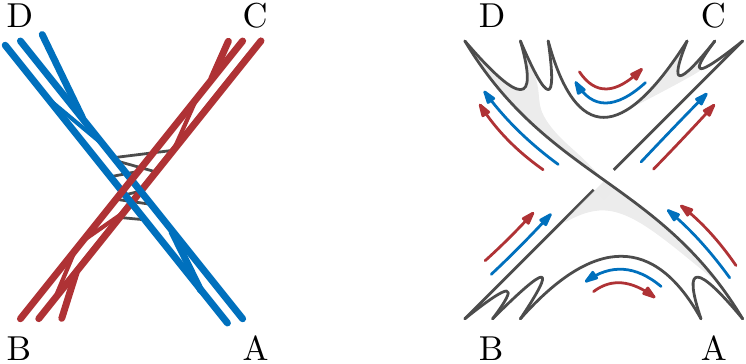}
	\caption{\label{fig:general}Left: If a singularity in the crossing domain were to develop, it would have to correspond to two beams of particles aligned in the directions of the external particles. Right: Edges on the sides of a planar diagram have definite components of the lightcone momenta in both $\BB{-}\CC$ and $\AA{-}\DD$ directions that forbid such singularities.}
\end{figure}

With this decomposition we can compute the imaginary parts of the on-shell conditions, which read
\be\label{eq:Imqe}
\Im(\hat{q}_e^2 - m_e^2) = \Im z  \sum_{\substack{a \in \AA\\ d \in \DD}} \sum_{\substack{b \in \BB\\ c \in \CC}} \frac{p_a^- p_b^- p_c^- p_d^-}{p_\AA^- p_\BB^-} \left(\frac{\theta_b \theta_c}{\theta_\BB} - \frac{1}{\,|z|^2} \frac{\theta_a \theta_d}{\theta_\AA}\right) f_{e,ad}\, f_{e,bc}.
\ee
In order to make the arguments as simple as possible we will require that each term in the parentheses is individually positive when $|z|^2 \geq 1$, that is
\be
\frac{\theta_a \theta_d}{\theta_\AA} < \frac{\theta_b \theta_c}{\theta_\BB}
\ee
for all choices of $a$, $b$, $c$, and $d$. If $\AA$ and/or $\DD$ consist of a single particle, and at the same time $\BB$ and/or $\CC$ consist of a single particle, these inequalities simply reduce to \eqref{eq:ratios}. In other cases, they further restrict the kinematic point defining the start of step $\RN{2}$. (To obtain such a point we can simply boost $z \to y z$ for a sufficiently large constant $y$, under which $\theta_b \theta_c / \theta_\BB \to y^2\, \theta_b \theta_c / \theta_\BB$.) As we have seen in the previous subsection, this additional requirement is not necessary for the proof of crossing symmetry in planar cases, but it will greatly simplify the analysis of the flow of energies in the diagram.

If there was a singularity, it would require that $f_{e,ad} f_{e,bc}=0$ for all the edges $e$ and all choices of $a$, $b$, $c$, and $d$ simultaneously. For a fixed $e$, this can only be true if all $f_{e,ad} = 0$, or all $f_{e,bc}=0$, with a possibility that both are true at the same time. Since the lightcone momenta take the form \eqref{eq:qepm}, such a singularity would have to correspond to two beams of particles, one purely aligned in the directions of $\BB{-}\CC$ and one in the directions of $\AA{-}\DD$, possibly exchanging massless states between them, as illustrated on the left panel of Fig.~\ref{fig:general}. Therefore, analyticity in the crossing domains hinges upon proving that such singularities do not appear, in either planar or non-planar diagrams.

We now specialize to planar Feynman diagrams specifically. Because of the arguments given in Sec.~\ref{sec:energy-flow}, such a singularity can never happen since the edges on the sides of the diagram have to have non-zero components in both directions. (By previous arguments, exceptions such as the one-vertex reducible diagrams are not singular in the crossing domains.) Precisely for such edges $e$, the factors $f_{e,ad}$ for all $a$ and $d$ must have the same sign and likewise the factors $f_{e,bc}$ for all $b$ and $c$ must have the same sign. This implies
\be
\Im(\hat{q}_e^2 - m_e^2) \neq 0,
\ee
which guarantees analyticity in the crossing domains, see the right panel of Fig.~\ref{fig:general}.

\vspace{-.5em}
\section{\label{sec:outlook}Outlook}

There are two natural avenues for future work. The first is optimizing for the size of the domains of analyticity. For example, in Sec.~\ref{sec:physical} it was sufficient for our purposes to show that there exists an infinitesimal neighborhood of the physical regions where scattering amplitudes are analytic, but we made no efforts to put bounds on the size of such neighborhoods. Similarly, in Sec.~\ref{sec:crossing} we showed that arranging the momenta in the lightcone according to Fig.~\ref{fig:lightcone} was sufficient for analyticity in the crossing domains, though it is not unlikely that some of these conditions can be relaxed.

The second question pertains to generalizations to non-planar scattering amplitudes. Since we already identified how potential singularities in the crossing domains would have to look like for \emph{any} diagram, planar or not (see the left panel of Fig.~\ref{fig:general}), one would have to prove that they cannot appear---at least in subregions of the crossing domains---without relying on the arguments of energy flows specific to planar diagrams  \footnote{One may ask if we can reasonably expect crossing symmetry to hold in perturbation theory for individual non-planar diagrams at all. At least in the cases proven non-perturbatively, we believe this will be the case because of the following basic argument. For a specific Feynman diagram we could construct a Lagrangian with $\E+n$ complex scalar fields (one for each internal and external line) with interaction vertices built in precisely such a way that to leading order in the couplings the whole scattering amplitude is dominated by a single Feynman diagram. To the extent that the perturbative approximation is valid, this suggests that the non-perturbative results carry over to individual Feynman diagrams regardless of planarity.}. We are aware of large classes of non-planar diagrams analytic in the crossing domains and no explicit counterexamples.

Let us finish by pointing out that the arguments we used in this work are much stronger than really necessary to prove analytic properties. Our derivation worked because certain propagators could not be put on-shell for any positive value of Schwinger parameters. By contrast, the necessary condition \emph{does} allow for every individual propagator to be on-shell, just not all of them being on-shell simultaneously. This leaves large room for improvement.

\vspace{1em}\noindent{\bf Acknowledgments.}
The author thanks Edward Witten for suggesting this problem and enlightening discussions, as well as Nima Arkani-Hamed, Simon Caron-Huot, Mathieu Giroux, Marcus Spradlin, and Andrew Tolley for useful comments and correspondence.
He gratefully acknowledges the funding provided by Frank and Peggy Taplin, as well as the grant DE-SC0009988 from the U.S. Department of Energy.

\appendix

\section{\label{sec:appendix}Clarifications About Fourier Transforms of Retarded Commutators}

In this appendix we clarify a misconception about the analyticity of scattering amplitudes perpetuated in the recent literature on effective field theories. The purpose of this discussion is to explain the difficulties in giving crossing symmetry a concrete physical interpretation at the non-perturbative level---even in the limited cases where it was proven---which in fact is the main motivation for reconsidering this problem in perturbation theory.

For simplicity we consider a four-particle scattering process in a scalar quantum field theory without massless particles. Following standard steps (see, e.g., \cite{Itzykson:1980rh}), the LSZ procedure allows us to write
\be\label{eq:Greens}
\G_{12 \to 34} - \G_{1\bar{3} \to \bar{2}4} \;=\; \int \d^\D x\, e^{i(p_2 - p_3)\cdot x}\, \langle -p_4 | \, [j_3^\dagger(x), j_2(-x)]\, | p_1\rangle,
\ee
where $j_a(x) = (\Box_x - M_a^2) \varphi_a(x)$ defines a current associated to the complex scalar field $\varphi_a$ and we ignore the overall normalization. Note that the combination $p_2^\mu {-} p_3^\mu$ is not the momentum transfer (in our conventions $t = (p_2 {+} p_3)^2$).
Whenever the integral can be defined, the left-hand side is a difference between two (amputated) off-shell Green's functions in the two crossing channels. Recall that in order to compute a scattering amplitude, one needs to take the on-shell limit, $p_i^2 \to M_i^2$, near the respective physical regions for both processes.

To establish crossing symmetry, one needs to prove that {\itshape(i)} the right-hand side of \eqref{eq:Greens} converges and is equal to zero in some region $R_4$ of the momentum space so that $\G_{12 \to 34} = \G_{1\bar{3} \to \bar{2}4}$ there, {\itshape(ii)} an analytic continuation from $R_4$ to the complex neighborhoods of the physical regions from the causal direction can be performed, and finally {\itshape(iii)} the two physical regions can be connected by a common domain of analyticity without leaving the on-shell kinematic space. The analogous question can be formulated for arbitrary multiplicity $n$.

The region $R_n$ can be determined by a careful examination of the conditions for polynomial boundedness of the integrand in \eqref{eq:Greens} and is called the \emph{primitive region} of analyticity (for each proper subset of particles $S$, it requires that either $\Im p_S^\mu$ is timelike or $\Im p_S^\mu = 0$ and $p_S^2 < \mathfrak{m}^2$ for any production threshold $\mathfrak{m}^2$) \cite{Steinmann1960a,Steinmann1960b,ruelle1961connection,doi:10.1063/1.1703695,araki1960properties}. Its derivation hinges on the assumptions of microcausality, locality, unitarity, and the mass gap. It has no support on-shell. At this stage of the computation, all the information about physics has been used and the remaining steps have to be performed using theorems in the analytic extension of $R_n$, which show that \emph{any} function---having physical origin or not---analytic in $R_n$ is also analytic in the extended region. The step {\itshape(ii)} can be performed for any multiplicity $n$, but with Green's functions represented as a boundary value of a \emph{single} analytic function only in the cases involving exactly two incoming or outgoing particles \cite{Bros:1964iho,Bros:1972jh}. The step {\itshape(iii)} proves to be much more complicated and has been only carried out for $n=4,5$ in massive theories for the crossing processes involving exactly two incoming particles \cite{Bros:1965kbd,Bros:1985gy}. Moreover, since the physical arguments employed in the step {\itshape(i)} only imply properties of off-shell Green's functions, and the later steps use complicated complex analysis theorems, it is difficult to associate concrete physical meaning to the results of such proofs.

What is the region of analyticity for general massive four-point scattering amplitudes? As mentioned above, in this case analyticity in the infinitesimal neighborhoods of the physical regions holds. The connection between them is given through the so-called \emph{asymptotic} crossing domains. The first one corresponds to the upper-half plane, $\Im s > 0$, of the center of mass energy $s$ with sufficiently large $|s|$ at any fixed squared momentum transfer $t = t_\star <0$. This high-energy small-angle limit can be used to connect the $s$-channel physical domain to the $u$-channel from the unphysical side, i.e., corresponding to the $-i\eps$ prescription. The remaining connection to the $t$-channel physical region can be achieved by an analogous asymptotic crossing domain with $s \leftrightarrow t$. The resulting region of analyticity is a subset of the one proven in this paper, though of course it holds non-perturbatively and does not require planarity but assumes the mass gap.

While the assumption of exchanged states being massive is quite central to the derivation (it guarantees that $R_n$ is non-empty), we believe that the aforementioned proofs can be extended to also include massless \emph{external} states, in the light of the more recent understanding of the infrared issues and the LSZ procedure for massless asymptotic states; see, e.g., \cite{Collins:2019ozc,Hannesdottir:2019rqq,He:2020ifr}.
Certain improvements on the domain of analyticity can be achieved if one assumes bounds on the masses involved in the process, e.g., that the external masses are sufficiently light compared to the internal ones, as well as sufficiently close to each other, in which case $R_n$ can have on-shell support and the analysis simplifies; see \cite{Bremermann:1958zz,Bogolyubov:104088,Hepp1964} for the $n=4$ case.

Following \cite{Meiman:1965gfz}, it has been recently suggested that complications associated with the analytic extension proofs can be avoided by taking a large-energy limit directly for the integral \eqref{eq:Greens}. If such a claim was indeed true, it would give a rather compelling physical explanation for crossing symmetry---even in the absence of more rigorous proofs---and hence deserves scrutiny.

Let us follow this chain of logic. We consider lightcone momenta for massive particles in the same notation as in \eqref{eq:4pt}. (This problem is typically stated in the Breit coordinate system, which introduces spurious square-root branching that we want to avoid in order to make the issue more transparent.) We take
\begin{gather}
p_1^\mu = \big(p_1^+,\,p_1^-,\, \vec{p}_1\big),\quad\;
p_2^\mu = \big(zp_2^+,\, \mfrac{1}{z}p_2^-,\, \vec{p}_2\big),\\
p_3^\mu = \big({-}zp_2^+,\, {-}\mfrac{1}{z}p_2^-,\, \vec{p}_3\big),\quad\;
p_4^\mu = \big({-}p_1^+,\, {-}p_1^-,\, \vec{p}_4\big),
\end{gather}
where $p_1^\pm > 0$ and $p_2^\pm > 0$.
In terms of the Mandelstam invariants we have
\begin{align}
s &= \big(p_1^+ + z p_2^+\big)\big(p_1^- + \mfrac{1}{z} p_2^-\big) - (\vec{p}_1 + \vec{p}_2)^2 \nn\\
&= z p_2^+ p_1^- \;+\; {\cal O}(z^0)
\end{align}
and
\be
t = -(\vec{p}_2 + \vec{p}_3)^2.
\ee
Therefore, in order to explore the high-energy region, where we expect to prove analyticity, we need to take $\Im z > 0$ and large $|z|$ with all the remaining variables fixed. In this limit, the exponential in the integrand of \eqref{eq:Greens} equals
\be\label{eq:exp}
e^{i(p_2 - p_3)\cdot x} = e^{i \left(z p_2^+ x^- + \frac{1}{z} p_2^- x^+ - (\vec{p}_2-\vec{p}_3)\cdot \vec{x}\right)} \approx e^{i z p_2^+ x^-},
\ee
where $x^\mu = (x^+, x^-, \vec{x})$. Using the fact that ${\cal G}_{12 \to 34}$ can be written in the same way as \eqref{eq:Greens}, except including the step function $\theta(x^0) = \theta(\frac{x^+ + x^-}{2})$ in the integrand, we find
\be\label{eq:G1234}
\lim_{z \to \infty} \G_{12 \to 34} \;\approx\; \int \d^\D x\, e^{i z p_2^+ x^-} \lim_{z \to \infty} \langle -p_4 | \,\theta(x^0)\, [j_3^\dagger(x), j_2(-x)]\, | p_1\rangle.
\ee
Under the assumption of microcausality, the commutator vanishes at spacelike separations and the integrand has support only in the future lightcone of $x^\mu$. (Inside the integral the retarded commutator $\theta(x^0)\, [j_3^\dagger(x), j_2(-x)]$ can be also expressed as the time-ordered product of the two currents \cite{Itzykson:1980rh}.) At this stage, one might conclude that the integral converges, and hence defines an analytic function, when the exponential is suppressed at infinity, i.e., when $\Re (i z p_2^+ x^-) < 0$ for large positive $x^-$. This is indeed the case when we consider $z$ in the upper-half plane, $\Im z > 0$. It would then appear that we have shown analyticity in $\Im s > 0$ for sufficiently large $|s|$.

This conclusion is at odds with the statement we made before, that Green's functions defined via the LSZ procedure do not in general converge on-shell. What went wrong? The first mistake was that we ignored the conditions on polynomial boundedness of the rest of the integrand of \eqref{eq:G1234} (for example, no constraints on $t$ or the masses were required). Correct analysis of this problem puts stringent conditions on the external kinematics, resulting in the primitive region $R_4$ described above. In particular, one can show that the Fourier transform never converges for positive $p^2_i = M_i^2$ unless the spectrum of the theory is truncated.

Even ignoring the first issue, the second mistake was to not be careful about the direction from which the high-energy limit has to be taken. For the exponential \eqref{eq:exp} to be suppressed when $x^\mu$ is in the future lightcone, we really need
\be
\Im (p_2^\mu - p_3^\mu) = 2 \Im z\, \big( p_2^+,\, -\mfrac{1}{\,|z|^2} p_2^-,\, \vec{0}\,\big)
\ee
to be future timelike or null. Since $p_2^- > 0$, the component along the negative axis of the lightcone is always negative, which means the above arguments only work when approaching the lightcone \emph{from the outside} (likewise, taking $p_2^- < 0$ would violate the on-shell conditions). Therefore, one cannot use any conclusions derived from such manipulations to \emph{directly} infer properties of scattering amplitudes.

\bibliography{references}

\end{document}